\documentclass[onecolumn,notitlepage,letterpaper,superscriptaddress,nofootinbib,longbibliography]{revtex4-2}

\usepackage{hyperref}
\usepackage{xurl}
\usepackage{graphicx}
\usepackage[font=footnotesize,labelfont=bf, width=0.7\textwidth]{caption}
\usepackage[bottom]{footmisc}
\usepackage{comment}

\usepackage[margin=0.6in]{geometry}
\usepackage{amsmath,amssymb}
 
\usepackage[bottom]{footmisc}

\newcommand{\beq} {\begin{equation}}
\newcommand{\eeq} {\end{equation}}

\begin{document}

\title{Pole inflation from extended metric-affine gravity} 

\author{Damianos Iosifidis}
\email{d.iosifidis@ssmeridionale.it}
\affiliation{Scuola Superiore Meridionale, Largo San Marcellino 10, 80138 Napoli, Italy}
\affiliation{INFN-Sezione di Napoli, Via Cintia, 80126 Napoli, Italy}

\author{Sotirios Karamitsos}
\email{sotirios.karamitsos@ut.ee}
\affiliation{Institute of Physics, University of Tartu, W.\ Ostwaldi 1, 50411 Tartu, Estonia}

 \begin{abstract}
We study inflation in the framework of extended metric-affine $F(R)$ gravity, where all even-parity quadratic invariants of torsion and non-metricity are included in the Lagrangian alongside the $F(R)$ term. The extended theory admits a scalar-tensor description with a non-canonical kinetic term featuring poles. As a result, the inflationary dynamics and predictions for observables of this model are insensitive to the specific form of $F(R)$, since they are dominated by the structure of the poles (order and residue). We analyze both a simplified version analytically and the full eleven-parameter theory, and we classify the models based on whether they feature second-order poles, whether they are free from ghosts, and whether they predict a sufficiently small tensor-to-scalar ratio. By relaxing the ghost-free requirement to only exclude ghosts near the pole (where inflation occurs), we demonstrate that we can significantly enlarge the set of viable models. We thus show that extended metric-affine $F(R)$ gravity can act as a robust framework for inflation, reproducing the attractor predictions for the spectral index and tensor-to-scalar ratio. 
\end{abstract}

\maketitle

\section{Introduction}

The formalism of Metric-Affine Gravity \cite{Hehl:1994ue} (MAG) offers a straightforward and elegant geometric  way to introduce new degrees of freedom that modify General Relativity (GR). These degrees of freedom can be used in a wide array of contexts, including the treatment of gravity as an EFT, coupling to
dark matter fermions, and early-time cosmology and inflation as well as late-time cosmology and dark energy. In standard GR, torsion and non-metricity vanish as the affine connection is \emph{a priori} taken to be the torsionless and metric-compatible Levi--Civita connection. However, in MAG, the linear connection is completely independent of the metric, which leads to an additional set of equations that must be solved to obtain it.

The new degrees of freedom provided by MAG have many applications, particularly in dark matter, dark energy, and inflation \cite{Andrei:2024vvy,Dyer:2024kvo,Bahamonde:2025eov,Iosifidis:2024bsq,Gialamas:2023flv,Gialamas:2024cit,Racioppi:2024zva,Dioguardi:2023jwa,Barker:2024dhb,Rigouzzo:2022yan,Rasanen:2018ihz}. In standard GR, such degrees of freedom can arise by modifying the Einstein--Hilbert action to the $F(R)$ action, which gives an additional scalar degree of freedom. In the MAG formalism, the situation is dramatically different; the vacuum pure $F(R)$ action only gives rise to a cosmological constant, in addition to GR \cite{Ferraris:1992dx,Sotiriou:2006qn}. Therefore, in order for MAG $F(R)$ theories to be dynamical without adding matter, additional geometric terms must supplement the gravitational sector.\footnote{Alternatively, one can supplement the Ricci scalar with quadratic curvature invariants. However, such modifications bring many unwanted  and problematic degrees of freedom and the healthy theories spectrum is very narrow \cite{Barker:2025xzd,Barker:2024ydb,Marzo:2021iok,Percacci:2020ddy}. } A straightforward approach, also motivated by an Effective Field Theory (EFT) viewpoint, is to add quadratic torsion and non-metricity invariants, which, unlike the $F(R)$ action, they are not projective invariant and consequently allow to excite additional degrees of freedom. As recently shown \cite{Iosifidis:2024ndl}, the theory\footnote{Namely the one whose gravitational Lagrangian consists of $F(R)$ along with the $11$ quadratic and parity even torsion and non-metricity invariants \cite{Pagani:2015ema,Iosifidis:2018zwo}.} is equivalent to a metric and torsionless scalar-tensor theory. The kinetic coupling of the resulting theory depends on the dimensionless parameters of the quadratic invariants only, whereas the scalar field potential depends solely on the functional form of $F(R)$.

It is well known that scalar-tensor theories featuring a scalar degree of freedom, which is in general coupled nonminimally to gravity or features a non-canonical kinetic term, have many applications in cosmology \cite{Fujii:2003pa,Faraoni:2004pi,Faulkner:2006ub, Quiros:2019ktw, Capozziello:2011et, Unnikrishnan:2012zu}. Such theories have attracted considerable attention, partly because many modified theories of gravity can be shown to have a scalar-tensor equivalent to them. Still, even within scalar-tensor theories, there are plenty of subclasses, and a particularly relevant one here is pole inflation \cite{Broy:2015qna, Terada:2016nqg, Kobayashi:2017qhk, Dias:2018pgj, Pallis:2022cnm}, which features asymptotically infinite kinetic terms, borne out of the $\alpha$-attractor models \cite{Kallosh:2013pby, Kallosh:2013lkr, Galante:2014ifa}. In general, every scalar-tensor theory featuring well-behaved couplings can be described by a single function (the canonical potential), but when discontinuities arise in the couplings, as is the case in pole inflation, the resulting equivalent canonical theory depends on the domain that inflation takes place in. This is precisely the type of model that the extended metric-affine $F(R)$ model is an example of.

Pole inflation is a particularly robust approach to model building, as the first-order phenomenological effects of inflation near the poles are entirely captured by the pole structure (their order and residue). Attractor theories of inflation, shown to lead to convergent predictions analogous to Starobinsky inflation, are very well in agreement with Planck \cite{Aghanim:2018eyx} and can be modified to match the potentially disruptive ACT results \cite{ACT:2025tim}. This means that the extended metric-affine $F(R)$ model, despite its cumbersome form, is, to first order, a second-order model of pole inflation, related to $\alpha$-attractors, and so can be studied with the powerful tools of pole inflation at our disposal. We can analytically study the observational viability of subregions within the parameter space by projecting down from the full model, but studying the full model itself will require some numerical techniques. The parameter space of this model features 11 parameters, but with some assumptions about naturalness (adopting an EFT viewpoint), we can randomly sample points in parameter space to determine which values for which parameters are favored and which are disfavored, as well as examine the viability of models with fixed values (e.g symmetric and general teleparallelism).

The structure of the paper is as follows: in Section~\ref{overview}, after going over some preliminaries, we provide the set-up that gives rise to the generic model for extended metric-affine $F(R)$ gravity, and we go over the basics of pole inflation and $\alpha$-attractors, demonstrating the effect of pole structure on the inflationary observables. In Section~\ref{toymodel}, we examine a simplified model that nonetheless features pole structure, and determine which parameter values are viable when taking into consideration the upper bound on gravitational wave production. In Section~\ref{fullmodel}, we consider the full extended metric-affine $F(R)$ model, identifying which regions of the parameter space are more favored by observations as well as considering a “compressed" version of the full model, featuring fewer parameters. Finally, we present our conclusions in Section~\ref{conclusion}. 

Throughout this article, we use natural units for which $\hbar = G = c =1$, and we employ the mostly plus signature for the metric. 

\section{Overview of MAG and pole inflation}
\label{overview}
\subsection{Review of Extended Metric-Affine $F(R)$ Gravity}

We begin by considering an $n$-dimensional manifold over which we introduce a metric $g_{\mu\nu}$ and an independent affine connection $\Gamma^{\lambda}{}_{\mu\nu}$ admitting a covariant derivative $\nabla_\mu$. On this non-Riemannian manifold, we define the curvature, torsion and non-metricity tensors as follows:
\begin{align}
R^{\mu}{}_{\nu\alpha\beta} &\equiv
2\partial_{[\alpha}\Gamma^{\mu} {}_{|\nu|\beta]}+2\Gamma^{\mu}{}_{\rho[\alpha}\Gamma^{\rho}{}_{|\nu|\beta]} \label{R},
\\
S_{\mu\nu}{}^{\lambda}&\equiv\Gamma^{\lambda}{}_{[\mu\nu]},
\\
Q_{\alpha\mu\nu} &\equiv- \nabla_{\alpha}g_{\mu\nu},
\end{align}
where the brackets denote antisymmetrization. The difference of the affine connection $\Gamma^{\lambda}{}_{\mu\nu}$ and the usual Levi-Civita connection defines the so-called distortion tensor \cite{Hehl:1994ue}
%schouten1954ricci
\begin{align}
N^{\lambda}{}_{\mu\nu} \equiv \Gamma^{\lambda}{}_{\mu\nu}-\widetilde{\Gamma}^{\lambda}{}_{\mu\nu}=
\frac{1}{2}g^{\alpha\lambda}(Q_{\mu\nu\alpha}+Q_{\nu\alpha\mu}-Q_{\alpha\mu\nu}) -g^{\alpha\lambda}(S_{\alpha\mu\nu}+S_{\alpha\nu\mu}-S_{\mu\nu\alpha}) \label{N}
\end{align}
where $\widetilde{\Gamma}^{\lambda}{}_{\mu\nu}$ is the usual Levi-Civita connection written solely in terms of the metric and its first derivatives. We can easily compute the torsion and non-metricity by using the distortion (see for instance \cite{Iosifidis:2018jwu}):
\beq
S_{\mu\nu\alpha}=N_{\alpha[\mu\nu]} \, , 
\qquad 
Q_{\nu\alpha\mu}=2 N_{(\alpha\mu)\nu} \label{QNSN}
\eeq
Out of the torsion and non-metricity, we can construct the vectors
\begin{align}
S_{\mu}  &\equiv S_{\mu\lambda}{}^{\lambda} \, , 
&
t_{\mu}  &\equiv \epsilon_{\mu\alpha\beta\gamma}S^{\alpha\beta\gamma} 
\\
Q_{\alpha} &\equiv Q_{\alpha\mu\nu}g^{\mu\nu}\, ,
&
q_{\nu} &\equiv Q_{\alpha\mu\nu}g^{\alpha\mu}
\end{align}
where, more precisely, $t^{\mu}$ is an axial vector (i.e. pseudovector).
From the curvature tensor we can form three contractions
\begin{align}
R_{\nu\beta}&=R^{\mu}{}_{ \nu\mu\beta}	
\\
\hat{R}_{\alpha\beta}&=R^{\mu}{}_{ \mu\alpha\beta}	
\\
\breve{R}^{\mu}{}_{ \beta}&=R^{\mu}{}_{ \nu\alpha\beta}	g^{\nu\alpha},
\end{align}
where $R_{\nu\beta}$ is the generalized Ricci tensor, $\hat{R}_{\alpha\beta}$ is the homothetic curvature, and $\breve{R}^{\mu}{}_{\beta}$ is the so-called co-Ricci tensor. The generalized scalar curvature $R$ is still uniquely defined, since
\beq
R \equiv g^{\mu\nu}R_{\mu\nu}=-g^{\mu\nu}\breve{R}_{\mu\nu} 
\eeq
while $g^{\mu\nu}\hat{R}_{\mu\nu}=0$.
 
A useful result is the following. Using \eqref{N}, we can split  any quantity into its Riemannian component (i.e. computed with respect to the Levi-Civita connection) and its  non-Riemannian component. For instance, inserting the connection decomposition \eqref{N} into the definition \eqref{R}, we obtain for the curvature tensor \footnote{Riemannian parts are denoted by $\widetilde{R}$, $\widetilde \nabla$,..., etc.} 
\beq
{R^\mu}_{\nu \alpha \beta} = \widetilde{R}^\mu_{\phantom{\mu} \nu \alpha \beta} + 2 \widetilde{\nabla}_{[\alpha} {N^\mu}_{|\nu|\beta]} + 2 {N^\mu}_{\lambda|\alpha} {N^\lambda}_{|\nu|\beta]} \,, \label{decomp}
\eeq
This expression is quite useful in metric-affine gravity. For instance, using this decomposition, the post-Riemannian expansion of the Ricci scalar is readily computed to be
\begin{align}
R \, =& \, \tilde{R} + \frac{1}{4}Q_{\alpha\mu\nu}Q^{\alpha\mu\nu}-\frac{1}{2}Q_{\alpha\mu\nu}Q^{\mu\nu\alpha}    -\frac{1}{4}Q_{\mu}Q^{\mu}+\frac{1}{2}Q_{\mu}q^{\mu}+S_{\mu\nu\alpha}S^{\mu\nu\alpha}-2S_{\mu\nu\alpha}S^{\alpha\mu\nu}-4S_{\mu}S^{\mu} \nonumber \\ &+2 Q_{\alpha\mu\nu}S^{\alpha\mu\nu}+2 S_{\mu}(q^{\mu}-Q^{\mu}) +\tilde{\nabla}_{\mu}(q^{\mu}-Q^{\mu}-4S^{\mu})
\end{align}

Having set up the preliminary concepts for metric-affine gravity, we turn our attention to the extended metric-affine $F(R)$ gravity model constructed in \cite{Iosifidis:2024ndl}.
This model in its most general form is constructed by taking the usual $F(R)$ Lagrangian and adding to it all parity even quadratic invariants of torsion and non-metricity, yielding \cite{Iosifidis:2024ndl}
 \begin{align}
	S
	=\frac{1}{2 \kappa}\int d^{4}x \sqrt{-g} \Big[ F(R)+\
	b_{1}S_{\alpha\mu\nu}S^{\alpha\mu\nu} +
	b_{2}S_{\alpha\mu\nu}S^{\mu\nu\alpha} +
	b_{3}S_{\mu}S^{\mu}+
	a_{1}Q_{\alpha\mu\nu}Q^{\alpha\mu\nu} +
	a_{2}Q_{\alpha\mu\nu}Q^{\mu\nu\alpha} \nonumber \\ +
	a_{3}Q_{\mu}Q^{\mu}+
	a_{4}q_{\mu}q^{\mu}+
	a_{5}Q_{\mu}q^{\mu} 
	+c_{1}Q_{\alpha\mu\nu}S^{\alpha\mu\nu}+
	c_{2}Q_{\mu}S^{\mu} +
	c_{3}q_{\mu}S^{\mu}
	\Big]  \label{S}
	\end{align}
The associated field equations are
\begin{align}
F' R_{(\mu\nu)}-\frac{f}{2}g_{\mu\nu}-\frac{\mathcal{L}_{2}}{2}g_{\mu\nu}+\frac{1}{\sqrt{-g}}(2S_{\alpha}-\nabla_{\alpha})\Big( \sqrt{-g}(W^{\alpha}{}_{(\mu\nu)}+\Pi^{\alpha}{}_{(\mu\nu)})\Big) +A_{(\mu\nu)}+B_{(\mu\nu)}+C_{(\mu\nu)}=\kappa T_{\mu\nu} \label{metricf}
\end{align}
 and
\begin{align}
& F'\left( \frac{Q_{\lambda}}{2}+2 S_{\lambda}\right)g^{\mu\nu}-F' (Q_{\lambda}^{\mu\nu}+2 S_{\lambda}^{\mu\nu})+F' \left( q^{\mu} -\frac{Q^{\mu}}{2}-2 S^{\mu}\right)\delta_{\lambda}^{\nu} \nonumber \\
+&\delta_{\lambda}^{\nu}\partial_{\mu}F'-g^{\mu\nu}\partial_{\lambda}F'+4 a_{1}Q^{\nu\mu}{}_{\lambda}+2 a_{2}(Q^{\mu\nu}{}_{\lambda}+Q_{\lambda}{}^{\mu\nu})+2 b_{1}S^{\mu\nu}{}_{\lambda}+2 b_{2}S_{\lambda}{}^{[\mu\nu]} \nonumber 
\\
+&c_{1}\Big( S^{\nu\mu}{}_{\lambda}-S_{\lambda}{}^{\nu\mu}+Q^{[\mu\nu]}{}_{\;\lambda}\Big)+\delta_{\lambda}^{\mu}\Big( 4 a_{3}Q^{\nu}+2 a_{5}q^{\nu}+2 c_{2}S^{\nu}\Big)+\delta_{\lambda}^{\nu}\Big(  a_{5}Q^{\mu}+2 a_{4}q^{\mu}+ c_{3}S^{\mu}\Big) \nonumber \\
+&g^{\mu\nu}\Big(a_{5} Q_{\lambda}+2 a_{4}q_{\lambda}+c_{3}S_{\lambda} \Big)+\Big( c_{2} Q^{[\mu}+ c_{3}q^{[\mu}+2 b_{3}S^{[\mu}\Big) \delta^{\nu]}_{\lambda} =0. \label{Gfieldeqs}
\end{align}
In the above equations, we have used the following definitions:
\begin{align}
W^{\alpha}_{\;(\mu\nu)} &\equiv 2 a_{1}Q^{\alpha}_{\;\mu\nu}+2 a_{2}Q_{(\mu\nu)}{}^{\alpha}+(2 a_{3}Q^{\alpha}+a_{5}q^{\alpha})g_{\mu\nu}+(2 a_{4}q_{(\mu} + a_{5}Q_{(\mu})\delta^{\alpha}_{\nu)}
\\
\Pi^{\alpha\mu\nu} &\equiv c_{1}S^{\alpha\mu\nu}+c_{2}g^{\mu\nu}S^{\alpha}+c_{3}g^{\alpha\mu}S^{\nu}
\end{align}
as well as 
\begin{align}
A_{\mu\nu}&\equiv a_{1}(Q_{\mu\alpha\beta}Q_{\nu}^{\alpha\beta}-2 Q_{\alpha\beta\mu}Q^{\alpha\beta}{}_{\nu})-a_{2}Q_{\alpha\beta(\mu}Q^{\beta\alpha}{}_{\nu)}
+a_{3}(Q_{\mu}Q_{\nu}-2 Q^{\alpha}Q_{\alpha\mu\nu})-a_{4}q_{\mu}q_{\nu}-a_{5}q^{\alpha}Q_{\alpha\mu\nu}
\\
B_{\mu\nu}&\equiv b_{1}(2S_{\nu\alpha\beta}S_{\mu}{}^{\alpha\beta}-S_{\alpha\beta\mu}S^{\alpha\beta}{}_{\nu})-b_{2}S_{\nu\alpha\beta}S_{\mu}{}^{\beta\alpha}+b_{3}S_{\mu}S_{\nu} 
\\
C_{\mu\nu} &\equiv\Pi_{\mu\alpha\beta}Q_{\nu}{}^{\alpha\beta}	-( c_{1}S_{\alpha\beta\nu}Q^{\alpha\beta}{}_{\mu}+c_{2}S^{\alpha}Q_{\alpha\mu\nu}+c_{3}S^{\alpha}Q_{\mu\nu\alpha})
\\
&\equiv c_{1}(Q_{\mu}{}^{\alpha\beta}S_{\nu\alpha\beta}-S_{\alpha\beta\mu}Q^{\alpha\beta}{}_{\nu})+c_{2}(S_{\mu}Q_{\nu}-S^{\alpha}Q_{\alpha\mu\nu})
\end{align}
After solving the connection field equations and consequently integrating out torsion and non-metricity from the metric field equations, it was shown in \cite{Iosifidis:2024ndl} that the above theory is equivalent on-shell to the theory captured by the following Lagrangian:
\beq
S=\frac{1}{2\kappa}\int d^n x \, \sqrt{-g}\Big[\Phi \tilde{R}-\frac{\omega(\Phi)}{\Phi}(\partial \Phi)^{2} - 2 V(\Phi) \Big] \label{EquivS}
\eeq
where the potential is given by solving $F_{,\phi} = \Phi$ and inverting to find $\phi(\Phi)$, giving:
\begin{align}
V(\Phi)= \frac{\Phi \, \phi(\Phi)-F(\phi(\Phi))}{2}.
\end{align}
The kinetic prefactor, on the other hand, is given by 
\begin{align}
    -\omega(\Phi) = \ &(n-1)\frac{\Phi^{2}}{P_{2}(\Phi)^{2}}\Big[ (C_{2}\Phi+D_{2})^{2}+ (C_{3}\Phi+D_{3})^{2}+n(C_{2}\Phi+D_{2})(C_{3}\Phi+D_{3})\Big]+ \nonumber \\
& (n-1)\frac{\Phi}{P_{2}(\Phi)}\Big[ (C_{2}-C_{3})\Phi +(D_{2}-D_{3}) \Big]+\left( \sum_{i=1}^{3}\sum_{j=1}^{3}(C_{i}\Phi+D_{i})\beta_{ij}(C_{j}\Phi+D_{j}) \right) \frac{\Phi}{P_{2}^{2}(\Phi)} \label{om}
\end{align}
where the constants $C_{i}, D_{i}$ are related to the parameters $a_{i}, b_{i}, c_{i}$ and $P_{2}(\Phi)$ is a quadratic polynomial in $\Phi$.  Therefore, the initial MAG theory (\ref{S}) with torsion and non-metricity is dynamically equivalent to the metric and torsionless scalar-tensor theory (\ref{EquivS}).

In the equivalent description, the expressions for torsion and non-metricity in terms of the extra scalar degree of freedom are found to be
\beq
S_{\mu\nu\alpha}=\frac{1}{P_{2}(\phi)}\Big( (C_{1}-C_{2})\phi +(D_{1}-D_{2})\Big) g_{\alpha[\mu}\partial_{\nu]}\phi
\eeq
with
\beq
S_{\mu}=\frac{(n+1)}{2 P_{2}(\phi)}\Big( (C_{1}-C_{2})\phi +(D_{1}-D_{2})\Big) \partial_{\mu}\phi
\eeq
and 
\beq
Q_{\nu\alpha\mu}=\frac{1}{P_{2}(\phi)}\Big[ (C_{1}\phi +D_{1})g_{\alpha\mu}\partial_{\nu}\phi +\Big( (C_{2}+C_{3})\phi +(D_{2}+D_{3})\Big) g_{\nu(\mu}\partial_{\alpha)}\phi \Big].
\eeq
The associated vectors are given by
\beq
Q_{\mu}=\frac{\tilde{C}_{1}\phi +\tilde{D}_{1}}{P_{2}(\phi)}\partial_{\mu}\phi \;\;, \;\; q_{\mu}=\frac{\tilde{C}_{2}\phi +\tilde{D}_{2}}{P_{2}(\phi)}\partial_{\mu}\phi
\eeq
where we have defined
\begin{align}
\tilde{C_{1}} & =n C_{1}+C_{2}+C_{3}, & \tilde{D_{1}}&=n D_{1}+D_{2}+D_{3},
\\
\tilde{C}_{2} & =C_{2}+\frac{(n+1)}{2}(C_{2}+C_{3}),& \tilde{D}_{2}& =D_{2}+\frac{(n+1)}{2}(D_{2}+D_{3}).
\end{align}
In general, we see that at the poles (i.e. when $P_{2}(\phi)=0$) both torsion and non-metricity blow up unless the degree of $P_{2}(\phi)$ reduces to one and the coefficients of the monomial match those of the numerators. We see for instance, quite interestingly, that when $P_{2}(\phi) \propto (C_{1}-C_{2})\phi+(D_{1}-D_{2})$, non-metricity blows up but torsion is regular.

The theory described in \eqref{om} belongs to the class of generalized Brans--Dicke theories with Brans--Dicke parameter $\omega(\Phi)$. Brans--Dicke theories \cite{Brans1961} have been extensively studied in a wide variety of contexts, from early-universe cosmology \cite{Paliathanasis:2015arj} and inflation \cite{Starobinsky:1994mh} to black holes \cite{Sotiriou:2011dz}.
When driving inflation, Brans--Dicke theories can be studied under the pole inflation formalism, since the presence of the $P_2(\Phi)$ polynomial term in the denominator of the kinetic term gives rise to poles, whose structure turns out to dominate the first-order expressions of the inflationary observables. Therefore, in the next subsection, we will provide an overview of pole inflation before we go on to use its techniques in studying the scalar-tensor Lagrangian admitted by the extended metric-affine $F(R)$ model.

\subsection{Review of pole inflation}

Pole inflation encompasses many models of inflation, and in particular the so-called attractor theories, including $\alpha$-attractors, named so because their predictions are insensitive to the overall shape of the potential. While there are many ways to motivate the appearance of poles in the Lagrangian, the predictions to first order are dependent solely on the pole structure in whose neighborhood inflation occurs.

The most common setup for pole inflation is that of a minimally coupled model that nonetheless features a non-canonical kinetic term, i.e. models with a Lagrangian of the form
\begin{align}\label{poleS}
S = \frac{1}{2\kappa}\int d^n \sqrt{-g} \left[ R -  k(\phi)  (\partial\phi)^2 - 2V(\phi)\right].
\end{align}
The kinetic prefactor $k(\phi)$ is further assumed to feature at least one pole at some point $\phi_p$, and we also assume that the potential does not feature any poles. It is possible to show that if the bulk of inflation occurs near $\phi_p$ (or any other pole), then the first-order predictions for (dimensionless\footnote{The amplitude $\Delta^2(k)$ is known as the dimensionless power spectrum, but that is due to normalization with respect to a reference scale, and so does depend on the details of the potential. In contrast, $n_s$ and $r$ are dimensionless without the need for introducing an arbitrary scale.}) quantities depend \emph{solely} on the features of the pole (the order and the residue) and not the potential. 

We note that the scalar-tensor equivalent to the extended metric-affine $F(R)$ model \eqref{EquivS} is not minimally coupled: it is a Brans--Dicke type action that belongs to the more general class of scalar-tensor theories, admitted by the action
\begin{align}\label{poleSb}
S = \frac{1}{2\kappa}\int d^n \sqrt{-g} \left[ f(\phi)  R -  k(\phi)  (\partial\phi)^2 - 2V(\phi)\right].
\end{align}
This action (said to be in the Jordan frame), can be recast such that the effects of the nonminimal coupling are contained in the kinetic term. Thus, it can be written in a form analogous to \eqref{poleS}. This is achieved via a conformal transformation \cite{Faraoni:1998qx}, that is, a field-dependent rescaling of the metric:
\begin{align}
g_{\mu\nu} \to \Omega(\phi)^2 g_{\mu\nu}.
\end{align}
By judiciously selecting $\Omega (\phi) = f(\phi)$, where the nonminimal coupling $f(\phi) = \phi$ for the metric equivalent of the extended $F(R)$ model, we can put the action into the following form
\begin{align}\label{poleS2}
S = \frac{1}{2\kappa} \int d^n \sqrt{-g} \left[ R -  G(\phi)  (\partial\phi)^2 - 2V(\phi)\right].
\end{align}
where the kinetic coupling now is
\begin{align}\label{Gdef}
G(\phi) = \frac{k(\phi )}{f(\phi )}+ \frac{n-1}{n-2} \left[\frac{f'(\phi )}{f(\phi )}\right]^2,
\end{align}
which is the single-field analogue to the field-space metric of multifield inflation.

For each pole individually, we now have two possibilities: the first is to assume that the potential is well-behaved around the kinetic pole, featuring no poles itself, and the second is to assume that the potential does feature a pole coinciding with the kinetic pole. We will first study the case in which there is no pole in the potential. We can (without loss of generality) shift the pole under consideration to $\phi = 0$. We can also assume that the potential has a negative gradient, as long as it does not feature a stationary point at the pole. We then keep the leading term in the Laurent series expansion of the kinetic term and the Taylor series expansion of the potential, giving us \cite{Terada:2016nqg}:
\begin{align}\label{linaction}
S = \frac{1}{2\kappa}\int d^4 x  \sqrt{-g} \left[ R - \frac{a_p}{ |\phi|^p} (\partial\phi)^2 - 2(V_0 - v_0 \phi) \right],
\end{align}
where $V_0 \equiv V(\phi = 0)$ and $v_0 = V'(\phi = 0)$, and we  have also set $n=4$. The kinetic term prefactor therefore sets the order $p$ and the residue~$\alpha_p$ of the pole.

In order to extract the observables, we proceed with the usual canonicalization procedure, defining a new canonical field $\varphi$ that has a standard kinetic term, achieved via
\begin{align}
\left(\frac{d\varphi}{d\phi}\right)^2 = \frac{a_p}{ |\phi|^p}.
\end{align}
This gives us
\begin{align}
\varphi =
\begin{cases} 
\frac{2 \sqrt{\alpha_p} |\phi|^{1 - \frac{p}{2}}}{p - 2} & (p \neq 2), \\[10pt]
\sqrt{\alpha_p} \ln |\phi| & (p = 2).
\end{cases}
\end{align} 
Since it is impossible for $\phi$ to cross a pole\footnote{Since its “mass" blows up at the pole or, equivalently, two points on different sides of a pole have an infinite field space distance between them.}, the canonical potential $U(\varphi) = V(\phi(\varphi))$ approaches the value $V_0$ as~$\varphi$ moves such that $\phi$ \emph{asymptotically} approaches the pole. In the canonical description, this corresponds either to a hilltop for $p<2$, or a plateau as the relation between $\varphi$ and $\phi$ “stretches" out the potential, leading to either a Starobinsky plateau ($p=2$) or an inverse hilltop plateau ($p>2$):
\begin{align}
U(\varphi) =
\begin{cases}
V_0 \left[ 1 - \left( \frac{p - 2}{2 \sqrt{\alpha_p}} \varphi \right)^{-\frac{2}{p - 2}} + \dots \right] & (p \neq 2), \\[10pt]
V_0 \left( 1 - e^{-\phi / \sqrt{\alpha_p}} + \dots \right) & (p = 2).
\end{cases}
\end{align}
We note that for inflation to occur, the potential needs to be such that it pushes the field \emph{away} from the pole (even if the kinetic term is non-canonical, the field will still always roll down the potential). Otherwise, the universe will accelerate eternally with no graceful exit. Therefore, if, at the pole, the potential has a positive gradient, inflation can only occur to the left, and if it has a negative gradient, it can only occur to the right. If the gradient is exactly zero at the pole, inflation can occur on both sides as long as the potential there is at a local maximum.

Writing the number of e-foldings through the usual slow-roll relation $\frac{dN}{d\varphi} = \frac{U(\varphi)}{U'(\varphi)}$:
\begin{align}
N =
\begin{cases}
\frac{\alpha_p}{v_0 (p - 1)} \left( \phi^{1 - p} - \phi_{\text{end}}^{1 - p} \right) & (p \neq 1), \\[10pt]
- \frac{\alpha_p}{v_0} \ln \frac{\phi}{\phi_{\text{end}}} & (p = 1).
\end{cases}
\end{align}
Using the standard definitions for $\epsilon$ and $\eta$, we arrive at the following predictions for $n_s$ and $r$:
\begin{align}
n_s = 1 -\frac{p}{(p-1)N},
\qquad
 r = \frac{8}{a_p} \left[ \frac{a_p}{(p-1)N} \right]^{\frac{p}{p-1}}.
\end{align}
These expressions assume that any poles in $V(\phi)$ do not coincide with the poles in $k(\phi)$, which is a reasonable assumption in most models. For a second-order pole, we have the usual attractor predictions
\begin{align}
n_s = 1 -\frac{2}{ N},
\qquad
 r = \frac{8 a_p}{N^2},
\end{align}
to second  order in $N^{-1}$. These predictions justify the “attractor" moniker applied to pole inflation: this has nothing to do with the usual slow-roll attractor solution in phase space (which corresponds to the lowest order in a series expansion in terms of the slow-roll parameters \cite{Liddle:1994dx}), but underlines that, as noted earlier, the overall shape of the potential does not matter when it comes to determining the observable quantities predicted by this theory. While the specific case of $p=2$ is somewhat disfavored by the recent ACT measurements for $N\approx 60$ \cite{Das:2025bws}, it is possible for these attractor models to survive, for example by modifying the details of the post-inflationary reheating scenario \cite{Haque:2025uri}, or introducing minute modifications to the attractor action \cite{He:2025bli}. 

We now turn our attention to the case in which the kinetic pole under consideration coincides with a pole in the potential. In this case, we must look at the leading order in the Laurent series expansion of the potential. The action then becomes 
\begin{align}\label{polepotaction}
S = \frac{1}{2\kappa} \int d^4 x  \sqrt{-g} \left[ R - \frac{a_p}{ |\phi|^p} (\partial\phi)^2 - \frac{2v_p}{|\phi|^s}   \right],
\end{align}
In this case, the observables become (for $p > 2$)
\begin{align}
n_s = 1 - \frac{p+s-2}{(p-2)N}, \qquad r = \frac{8s}{(p-2)N}.
\end{align}
For $p = 2$ instead, we find ourselves in an interesting situation. Canonicalizing the field as usual, we find ourselves in power-law inflation, ruled out by Planck measurements but possibly compatible with ACT \cite{Das:2025bws}. In this case, power-law inflation returns:
\begin{align}\label{powerlaw}
n_s = 1 - \frac{s^2}{\alpha_p}, \qquad r = \frac{8s^2}{\alpha_p}.
\end{align}
This provides the usual power-law consistency relation between $n_s$ and $r$.

\section{The $Q^2 + S^2$ model}
\label{toymodel}

We will now consider a toy model which is essentially a simplified version of the $F(R)$ extended MAG model. Instead of the full set of parity quadratic invariants, we will construct the action out of the Weyl vector and the torsion vector, giving rise to a simpler action that nonetheless includes both effects of torsion and non-metricity. We will then study the pole structure of the equivalent metric model, which we will use to impose restrictions on the parameter space coming from both the requirement that the theory is ghost-free as well as upper bounds on the production of gravitational waves.  

\subsection{Scalar-tensor description}

The theory we will consider is given by
\begin{align}
S = \frac{1}{2\kappa} \int d^n x \, \sqrt{-g} \left[ F(R) + \beta Q_\mu q^\mu + \gamma S_\mu S^\mu\right]
\end{align}
where $Q_\mu$ is the Weyl vector (which comes from non-metricity) and $S_\mu$ is the torsion vector. Varying the action as usual with respect to the metric and the connection gives us the field equations. After some algebra, it follows that $Q_\mu = \lambda S_\mu$ and in turn both are sourced by $F'(R)$. Then, with the usual field redefinition $\phi = F'(R)$, it is possible to show that the theory propagates an additional degree of freedom, and is in fact equivalent to a specific metric and torsionless scalar-tensor model: a Brans-Dicke theory with non-minimal coupling $f(\phi) = \phi$ as usual, and kinetic coupling $k(\phi)=\omega(\phi)/\phi$, given by \cite{Iosifidis:2024ndl}
\begin{align}
k(\phi) = -\frac{(n-1) }{(n-2)}  \left[(A \phi +B)^2-\frac{\gamma ^2}{n^2}+\frac{(n-1) \phi  \left(\beta  \lambda ^2+\gamma \right)}{n-2}\right] \frac{1}{  \phi  (A \phi +B)^2 },
\end{align}
where the constants $A$, $B$, and $\lambda$ stand for
\begin{align}
A &\equiv \frac{\gamma  (n-2)}{2}    \left(\frac{4}{\gamma  (n-1)}-\frac{n-1}{4 \beta  n^2}\right),
\\
B &\equiv -\frac{2 \gamma }{n^2},
\\
\lambda &\equiv \frac{n-1}{n} \frac{\gamma  }{4 \beta  }.
\end{align}
We can also write the potential, which we find is, as usual
\begin{align}
V(\phi) = \frac{1}{2}\left[ \phi \, \Phi (\phi) - F(\Phi(\phi)) \right]
\end{align}
where we must evaluate $F'(\Phi) = \phi$ first before solving for $\Phi$ in order to find the explicit functional form of~$\Phi(\phi)$. It is important to note that it has no dependence on $\beta$ or $\gamma$, since it is only implicitly affected by the non-metricity or the torsion.

In order to ensure that there are no ghosts, we must demand that $G(\phi)$ as defined in \eqref{Gdef} remains positive in the domain where the field evolves. Restricting ourselves to $n=4$, we find
\begin{align}\label{kterm}
G(\phi) = \frac{3}{2\phi^2}\left[1 - \frac{(\phi - \Delta_-) (\phi - \Delta_+)}{(\phi-\Delta)^2}\right]
\end{align}
where we have defined the following constants:
\begin{align}
\Delta &\equiv \frac{24 \beta  \gamma }{256 \beta -9 \gamma }
\\
\Delta_{\pm} &\equiv 
-\frac{12 \beta  \gamma  \left(1792 \beta + 99 \gamma \pm \sqrt{3997696 \beta ^2+299520 \beta  \gamma +10773 \gamma ^2}\right)}{(256 \beta -9 \gamma )^2}.
\end{align}                                             
Crucially, the sign of \eqref{kterm} is given by the sign of the following expression: 
\begin{align}
\delta(\phi)
&=
 32 \beta ^2 \gamma ^2-768 \beta ^2 \gamma  \phi -27 \beta  \gamma ^2 \phi  ,
\end{align}
which can be found by writing $k(\phi)$ in terms of $\beta$ and $\gamma$ and taking out an always positive denominator. 

\subsection{Pole structure and stationary points of $F(R)$ potentials}

As usual for single-field inflation, $n$ poles give rise to $n+1$ domains in which the field can evolve \cite{Karamitsos:2019vor}. In this case, the field can live in one of three domains, which correspond to the following intervals:
\begin{align}
I_- &= (-\infty, 0) \, ,
\\
I_0 &= (0, \Delta) \, ,
\\
I_+ &= (\Delta, \infty) \, .
\end{align}
Depending on the shape of the potential, inflation may be viable in any of the domains. However, near $\phi=0$, we may not find ourselves in an attractor regime since the kinetic pole coincides with the potential pole. We  write down the Einstein frame potential in terms of~$\phi$ (i.e. without canonicalizing the field) as
\begin{align}
U(\phi) = \frac{1}{2}\frac{ \phi \, \Phi (\phi) - F(\Phi(\phi))}{\phi^2}.
\end{align}
We remind that $\Phi(\phi)$ is found by solving the algebraic equation $\phi = F'(\Phi)$. Generically this potential will feature a pole at $\phi = 0$, and so, irrespective of its pole structure, the predictions are given by \eqref{powerlaw}. 

Furthermore, the side of the pole in which inflation can occur is determined by the side that the field rolls away from the pole. This is controlled by the derivative of the Einstein frame potential, $\frac{dU}{d\varphi} = \frac{dU}{d\phi} \frac{d\phi}{d\varphi}$. Assuming that there are no ghosts near the pole, it suffices to study the gradient of $U(\phi)$ in order to examine which direction the field can roll.

We can further seek a stationary point for the potential. Therefore, we consider $U'(\phi) = 0$, a condition which can be written as
\begin{align}
\frac{-\phi  \Phi '(\phi ) F'(\Phi (\phi ))+2 F(\Phi (\phi ))+\phi ^2 \Phi '(\phi )-\phi  \Phi (\phi )}{2 \phi ^3} = 0.
\end{align}
For $\phi\ne 0$ and using the definition $\phi = F'(\Phi)$ we have
\begin{align}
F(\Phi(\phi)) = \frac{\phi \Phi(\phi)}{2},
\end{align}
from which we obtain
\begin{align}
F(\Phi) = \frac{F'(\Phi) \Phi }{2}.
\end{align}
Note that this too is an algebraic equation: the form of $F(\Phi)$ is already fixed, and solving it returns the values of $\Phi$ where the potential $U(\phi)$ has stationary points\footnote{Solving it as a differential equation returns the only form of $F(R)$ in which the condition is satisfied for \emph{every} point, corresponding to an exactly flat canonical potential.} (which can be converted to values of $\phi$ as usual). For instance, for a simple $R+R^2$ theory, we find that the only stationary point is at $\Phi = 0$, or $\phi = 1$. Should this potential correspond to one of the poles, then inflation can occur on both sides of the pole. In any other case, if inflation is viable on one side of any pole, it will not be viable on the other. In general, if at least one pole and domain can give rise to good inflation, we can select an appropriate $F(R)$ such that the field rolls in the desired direction.  

\subsection{No-ghost constraint}

\begin{figure}[!ht]
\centering
\includegraphics[scale=0.35]{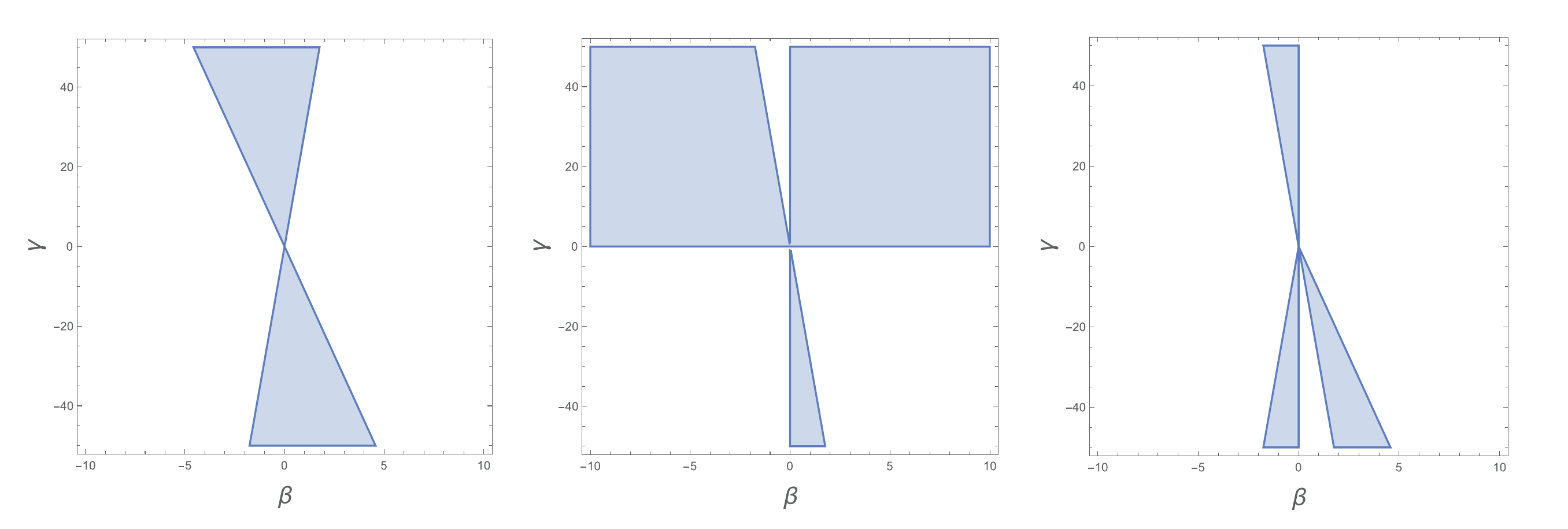}
\caption{Ghost-free regions in the parameter space for the $Q^2 + S^2$ model for inflation in $I_{0}$, $I_-$, and $I_+$, respectively.}\label{fig1} 
\end{figure}

We can now examine the ghost-free condition, which distinguishes between allowed and disallowed values for $\beta$ and $\gamma$ through $\delta(\phi)$ (which must remain positive throughout the domain):

\begin{itemize}
\item {\bf $I_0$ (within the poles).}  The discriminant $\delta(\phi)$ is a linear function, and $\delta(0) = 32 \beta^2 \gamma^2>0$. Therefore, it  suffices to require that $\delta(\Delta)>0$ to ensure that it is positive for all values of $\phi$ between the poles. This leads us to the following condition on $\beta$ and $\gamma$:
\begin{align}
\frac{1280 \beta +117 \gamma }{256 \beta -9 \gamma } < 0.
\end{align}

\item {\bf $I_-$ (before the poles).} Since $\delta(0) >0$ as noted above, in order to ensure that $\delta(\phi)$ remains positive for negative $\phi$, all we have to do is ensure a negative gradient, which leads us to the following condition on $\beta$ and $\gamma$:
\begin{align}
\beta  \gamma    (768 \beta +27 \gamma )>0.
\end{align}

\item {\bf $I_+$ (after the poles).} Here, the constraint is a combination of the $I_0$ constraint (which comes from ensuring $\delta(\Delta)>0$) and the inverse of the $I_-$ constraint (which comes from requiring a \emph{positive} gradient):
\begin{align}
\frac{1280 \beta +117 \gamma }{256 \beta -9 \gamma } &< 0,
\\
\beta  \gamma    (768 \beta +27 \gamma )&<0.
\end{align}
\end{itemize}
The allowed regions are visualized in Fig.~\ref{fig1}.
 
\subsection{Observational constraints}

\begin{figure}[!ht]
\centering
\includegraphics[scale=0.45]{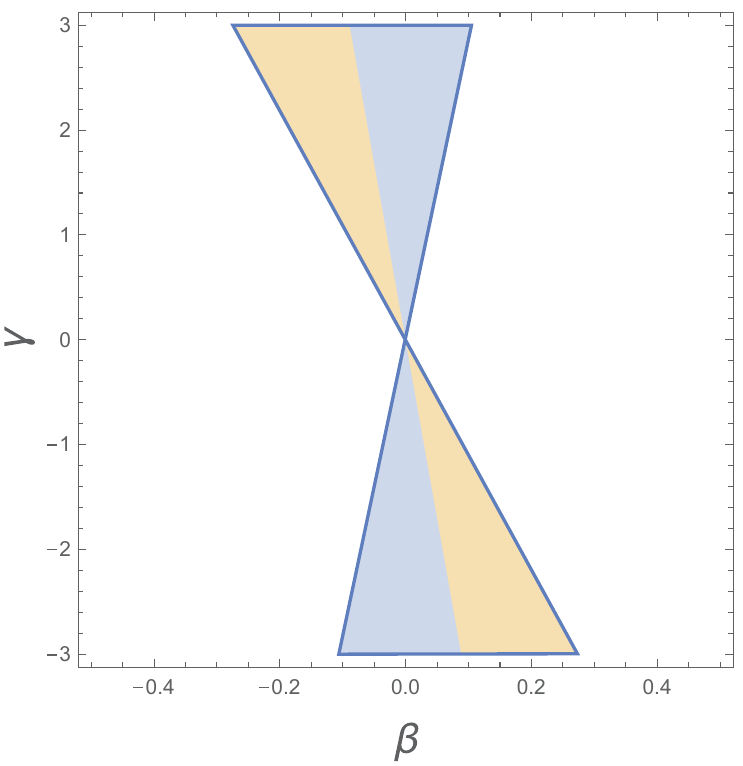}
\caption{Allowed regions in the parameter space for the $Q^2 + S^2$ model for inflation around the pole at $\phi  = \Delta$.  The blue and orange regions together correspond to a ghost-free theory and the orange  region alone satisfies the observational constraints on $r$.} \label{fig2}
\end{figure}

In pole inflation, the predictions for $n_s$ and $r$ depend solely on the nature of the pole. Instead of the domain we are in ($I_-, I_0$, or $I_+$), which sets the condition that avoids tachyonic modes, the predictions depend on the residue. First, we examine the pole at $\phi = \Delta$, for which we find
\begin{align}
r = \frac{24 }{N^2} \frac{(\Delta-\Delta_-)(\Delta_+ -\Delta)}{\Delta^2}.
\end{align}
After some tedious algebra, we find that, in terms of $\beta$ and $\gamma$, the prediction is
\begin{align}
r = -\frac{24}{N^2} \frac{(1280 \beta +117 \gamma )}{(256 \beta -9 \gamma ) }.
\end{align}
Note that $r>0$ is guaranteed by the no-ghost constraints in $I_0$ and $I_+$, which are the domains adjacent to $\phi = \Delta$. This further constrains the values of $\beta$ and $\gamma$ such that $r \lesssim r_{\rm max} = 0.036$ at 95\% CL \cite{Tristram:2021tvh}, or
 \begin{align}
 0 \le - \frac{(1280 \beta +117 \gamma )}{(256 \beta -9 \gamma ) } \lesssim  \frac{N^2 r_{\rm max}}{24}.
\end{align}
For $N = 60$, this places a further constraint on $\beta$ and $\gamma$ (note that unlike the ghost-free constraint, the observational constraint prevents $\beta$ and $\gamma$ from both being positive), shown in Fig.~\ref{fig2}. For the recent ACT observations \cite{ACT:2025tim}, this implies a value of $N \sim 70$ if inflation occurs near the $\phi = \Delta$ pole, which is consistent with $T_{\rm reh} \lesssim 10^{10} \ {\rm GeV}$ \cite{Drees:2025ngb}. For the $\phi = 0$ pole, where we expect a singularity in the potential as well, we find ourselves in power-law inflation with a constant $\epsilon$, leading to the usual consistency relation $r = 8(1-n_s)/3$.

\section{The extended MAG $F(R)$ model}
\label{fullmodel}

\subsection{Scalar-tensor description}

There are 11 even parity quadratic invariants in torsion and non-metricity, leading to a more complicated model, given by
\begin{align}
S = \frac{1}{2\kappa} \int d^4 x \, \sqrt{-g} \, 
\Big[ 
F(R)
&+b_1 S_{\alpha\mu\nu}  S^{\alpha\mu\nu}
+b_2  S_{\alpha\mu\nu}  S^{\mu\nu \alpha}
+b_3 S_\mu S^\mu
+a_1 Q_{\alpha\mu\nu}  Q^{\alpha\mu\nu}
+a_2 Q_{\alpha\mu\nu}  Q^{\mu\nu \alpha}
\nonumber\\
&+a_3 Q_\mu Q^\mu+a_4 q_\mu q^\mu+a_5 q_\mu Q^\mu
+c_1 Q_{\alpha\mu\nu}  S^{\alpha\mu\nu}
+c_2 Q_{\mu}  S^{\mu}
+c_3 q_{\mu}  S^{\mu}
\Big]
 \end{align}
Varying with respect to the connection and taking the independent contractions, we find a system for $Q_\mu$, $S_\mu$, and $q_\mu$ in terms of combinations of $a_i$, $b_i$, and $c_i$. 

As with the model studied in the previous section, we may cast it in terms of a Brans--Dicke type model. The nonminimal coupling is $f(\phi) = \phi$ just as before, and the potential will generically have a pole at $\phi = 0$. More importantly, using the kinetic prefactor $\omega(\phi)/\phi$ in the Jordan frame, we can (after some cumbersome calculations), determine the prefactor in the Einstein frame:
\begin{align}\label{Gcomp}
G(\phi)= \frac{3}{2\phi^2}\left[ 1+ \frac{\mu_1 \phi + \mu_2 \phi^2 + \mu_3 \phi^3 + \mu_4 \phi^4}{ (A_0 \phi^2 +B_0\phi +C_0)^2} \right]
\end{align}
where the explicit forms of $\mu_i$ can be found in Appendix~\ref{expl}, defined in terms of multiple constants that are in turn defined in terms of other constants, all the way to the basic parameters $a_i$, $b_i$, and $c_i$. 

Writing $\mu_i$ explicitly is a fool's errand: the resulting expressions are a fifth order polynomial in $a_i$, $b_i$, and $c_i$ with a total of 5652 terms! $A_0$, $B_0$ and $C_0$ are much more manageable (with $\mathcal{O}(10)$ terms), but their explicit forms do not offer much in the way of insight. Therefore, we will instead numerically work with the redefined constants in studying the pole structure of the theory.

\subsection{Second order, no-ghost, and observational constraints}

We note that, similar to the simple model studied in the previous section, the choice of parameters can be classified according to whether they give rise to ghosts, and whether the tensor-to-scalar ratio is within observational bounds. However, in the general case, there is another condition related to the order of the poles: in order to be able to study inflation other than power-law inflation, we must ensure that the theory actually can feature an order-two pole. This is not automatically satisfied in the general case: in fact, it occurs when the denominator of \eqref{Gcomp} has two roots, i.e. when $\Delta = B_0 ^2 - 4 A_0 C_0 > 0 $, giving rise to the poles at $\phi = \phi_\pm$, given by
\begin{align}
\phi_\pm = \frac{-B_0 \pm \sqrt{B_0^2-4 A_0 C_0}  }{2 A_0}.
\end{align}
Furthermore, there is always a pole at $\phi = 0$ as usual, which as mentioned in the previous section, leads to power-law inflation (the consistency relation remains the same).  If $B_0 ^2 = 4 A_0 C_0$, instead of three second-order poles (one at zero and the other two at the roots of the denominator), we have a second-order pole at zero and a fourth-order pole at the root of the denominator. We will ignore this case, as $p = 4$ significantly clashes with observations, while the pole at 0 is already captured by power-law inflation. 
 
Apart from the second-order pole condition, we must return to the no-ghost condition to ensure that the kinetic prefactor in the Einstein frame $G(\phi)$ is positive everywhere (again corresponding to a positive-definite field space metric) in the domain of interest. To begin with, we can impose a stronger condition, which is that the kinetic term is positive everywhere, ensuring that the theory does not feature ghosts in any domain. We look at the following expression that has the same sign as $G(\phi)$ (similar to before, we have taken out an always positive denominator):
\begin{align}
\delta(\phi) = C_0^2 + \left(2 B_0 C_0 + \mu_1\right) \phi + \left(B_0^2 + 2 A_0 C_0 + \mu_2\right) \phi^2 + \left(2 A_0 B_0 + \mu_3\right) \phi^3 + \left(A_0^2 + \mu_4\right) \phi^4.
\end{align}
In order for $\delta (\phi) > 0 $ to hold for all $\phi$, the polynomial must have no real roots. The condition for this to occur is explained in Appendix~\ref{appB}: in brief, the discriminant must be positive such that all roots are complex, and at least one of two additional polynomials (defined in terms of the coefficients) must be positive. We also require that the leading coefficient is positive, such that if there are no roots, the term remains positive everywhere.

Finally, for three distinct poles (two of which are second-order), we find that the tensor-to-scalar ratio at the poles is given by $r=12\alpha/N^2$ as usual, where the residues are found to be
\begin{align}
\alpha_{\pm} = \frac{6 A_0^2 \left( s_\pm \mu_1 + s_\pm ^2 \mu_2^2 + s_\pm ^3 \mu_3^3 + s_\pm ^4 \mu_4^4 \right)}{\left(B_0^2-4 A_0 C_0\right) a_\pm^2 }
\end{align}
where
\begin{align}
s_\pm \equiv \frac{ \phi_\pm}{2A_0},
\end{align}
depending on whether we are considering the $\phi_+$ or $\phi_-$ pole. The value of the residue will determine whether a particular value for the parameters can lead to an observationally viable theory.  
 
In summary, any choice of parameters can be checked for three conditions: 
\begin{enumerate}
\item \emph{the second-order pole condition}, which ensures that there is a second-order pole \emph{not} at zero, thus departing from the predictions of power-law inflation,

\item \emph{the no-ghost condition}, which ensures that the kinetic term does not become phantom by picking up the wrong sign to at least one of the domains on either side of the pole,

\item \emph{the observational viability condition}, which ensures  the tensor-to-scalar ratio (for some number of e-folds) near at least one  second-order pole is within the measured bounds\footnote{The spectral index is of course also part of the observables, but as it is not affected by the residue, it is not relevant in discriminating between favored and unfavored values for the parameters of the theory.}.
\end{enumerate}

The no-ghost condition is reasonable to request of a healthy theory. However, the existence of ghosts in some domain does not completely rule out inflation near the poles. As the field evolves, it is only necessary that there are no ghosts until inflation ends, which occurs roughly at $\epsilon = 1$. The approximation of $\epsilon = \phi^2/(2 \alpha_p)$ fails at that point as it is no longer near the pole and the leading term no longer strongly dominates in the Laurent series expansion, and after which post-inflationary effects may dominate the action, ridding it of its pole structure. Therefore, we can modify the no-ghost condition to the \emph{near no-ghost condition}:

\begin{enumerate}
\setcounter{enumi}{3} 
\item \emph{the near no-ghost condition}, which ensures that the kinetic term remains positive on at least one side of the pole and within the domain in which inflation happens (for which $\epsilon<1$). 
\end{enumerate}

We note that the second-order pole condition is global, whereas the rest of the conditions depend on which pole we consider. Therefore, to examine whether a particular choice of parameters can lead to a theory, we check whether the conditions are satisfied for at least one of the second-order poles.
%given our agnostic approach to the rest of the model (in particular since we do not fix the value of $F(R)$).

\subsection{Analytic approach: parameter space reduction}

\begin{figure} 
\centering
\includegraphics[scale=0.4]{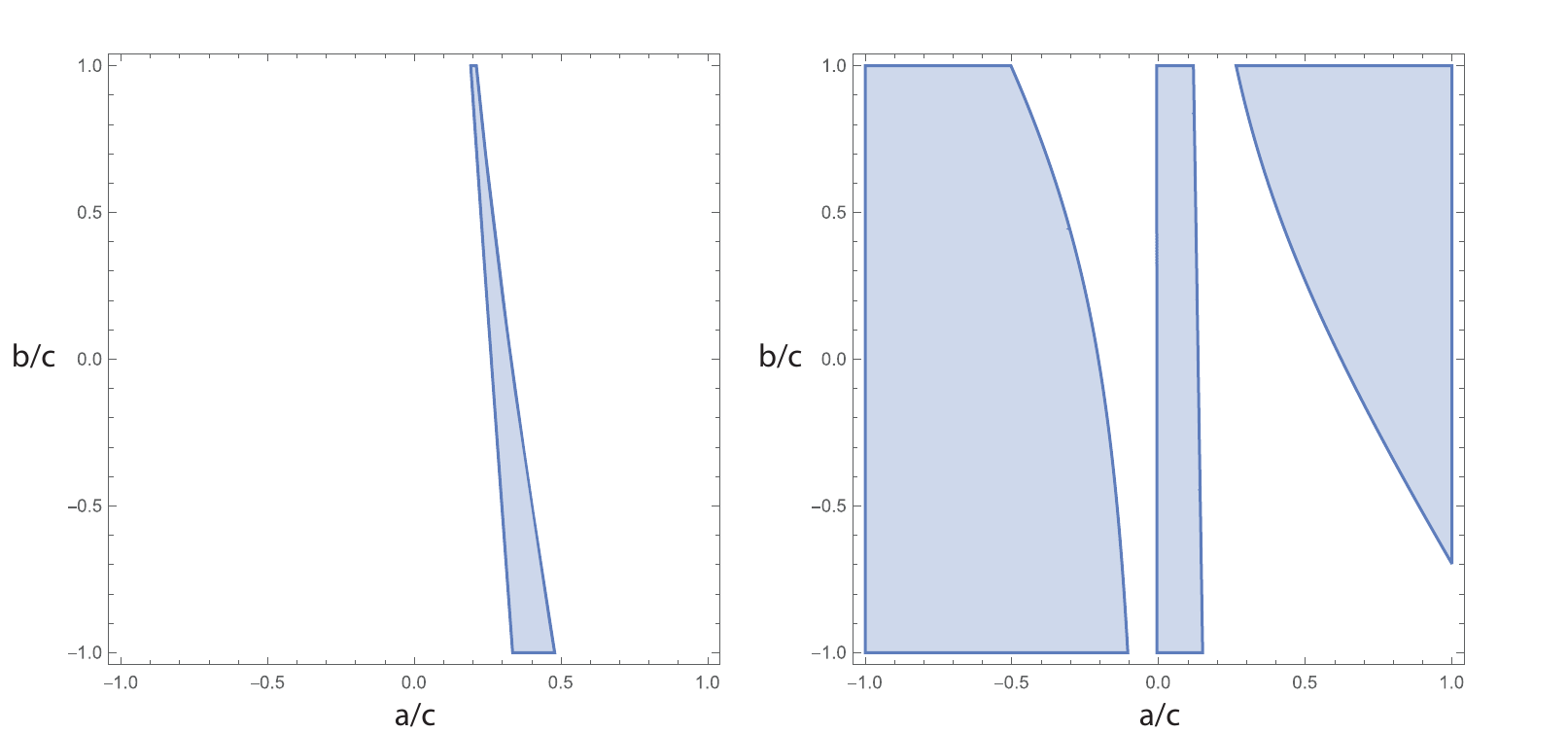}
%\qquad
%\includegraphics[scale=0.4]{poleQplotc1.pdf}
\caption{
Allowed regions in the parametrically reduced model. {\bf Left:} the no-ghost region. {\bf Right:} the second-order pole region, with the overlap specifying viable regions.}\label{fig3}
\end{figure}

Despite the analysis in the previous subsection in terms of the redefined parameters, the full theory in terms of the original coefficients $a_i$, $b_i$, and $c_i$ is impossible to attack intuitively. Instead, we can interpret the parameters of the theory as originating from some perturbations to the non-metricity and the torsion. This is reflected in the notation: the $a_i$ parametrize the effects of non-metricity, the $b_i$ parametrize the effects of torsion, and the $c_i$ parametrize the effects arising from a coupling between the two. Therefore, as perturbations of different sources, we may reduce the number of parameters to three by identifying:
\begin{align}
a &\equiv a_1 = a_2 = a_3 = a_4 = a_5,
\\
b &\equiv b_1 = b_2 = b_3,
\\
c &\equiv c_1 = c_2 = c_3.
\end{align}
In this case, the pole condition becomes
\begin{align}
\frac{7744 \bar a^2 \bar b^2}{81}+\frac{520960 \bar a^3 \bar b}{81}-\frac{24640 \bar a^2 \bar b}{27}+\frac{774400 \bar a^4}{81}-\frac{140800 \bar a^3}{27}-\frac{19360 \bar a^2}{27}-\frac{176 \bar a \bar b}{9}+160 \bar a +1 >0,
\end{align}
where we define the independent parameters $\bar a \equiv a/c$ and $\bar b = b/c$ (the no-ghost condition involves a 25th order polynomial in $\bar a$ and $\bar b$ and is too cumbersome to reproduce here). The distinct second-order poles are given by 
\begin{align} \textstyle
\phi_\pm = \frac{\pm\sqrt{352 {\bar a}^2 \left(22 {\bar b}^2-210 {\bar b}-165\right)+14080 {\bar a}^3 (37 {\bar b}-30)+774400 {\bar a}^4-144 {\bar a} (11 b-90)+81}-880 {\bar a}^2+{\bar a} (240-88 {\bar b})+9}{416 {\bar a}-36},
\end{align}
along with the usual pole at zero.

\begin{figure}[t]
\centering
\includegraphics[scale=0.5]{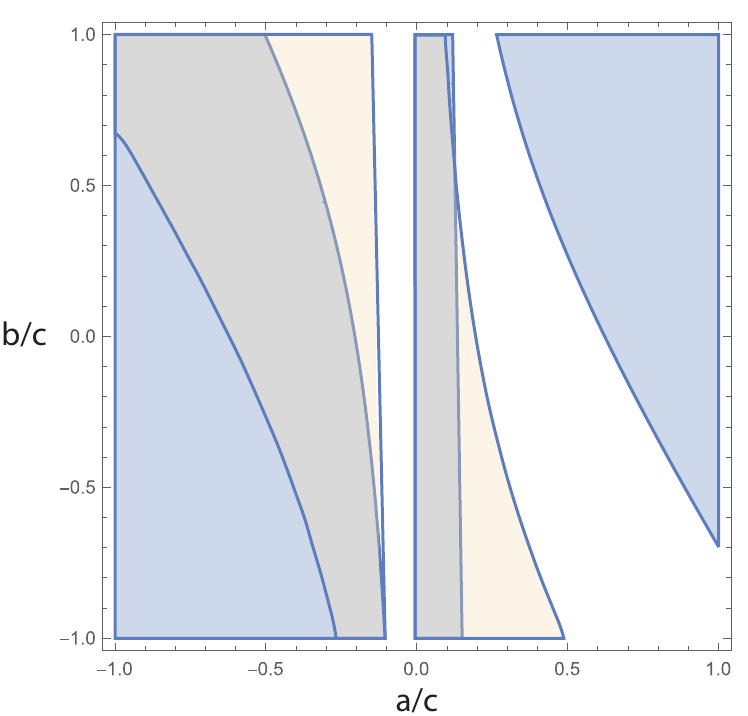}
\caption{Allowed regions in the parametrically reduced model with the no-ghost condition relaxed to allow ghosts as long as they do not occur between the pole and the end of inflation. The orange region corresponds to the relaxed no-ghost condition and the blue region corresponds to the second-order pole region.}\label{fig4}
\end{figure}

The residue of the resultant non-zero poles is small enough to lead to observationally viable inflation for both non-zero second-order poles, but we find that there is no overlap between the areas where the no-ghost condition and the separate pole condition is satisfied, as shown in Fig.~\ref{fig3}. This indicates that such a model will necessarily feature ghosts as the field moves further away from the pole. However, as noted in the previous subsection, if we relax the no-ghost condition to only disallow ghosts close to the pole, we may find that the allowed regions are wider. Indeed, if we impose the condition that the kinetic term should stay positive between the pole and the end of inflation (on at least one side), we find that there is a region where we can have both a second-order pole alongside with no ghosts during inflation, as shown in Fig~\ref{fig4}. 

\subsection{Stochastic approach: parameter sensitivity}

In the previous section, we examined conditions for the parameters of the simplified model that lead to a well-behaved theory. However, studying the full model is a much more demanding task, and visualizing the results of the study even more so, given that the parameter space is 11-dimensional. Therefore, we will adopt a stochastic approach in studying the parameter space of the full model numerically. 

Whether a particular choice of parameters satisfies the distinct pole condition and the no-ghost condition is a binary test. Satisfying the observational bound on the tensor-to-scalar ratio is not strictly a binary test (since the predicted value for $r$ corresponds to a certain $p$-value given the observational bounds), but we can consider a fixed confidence level (e.g. the $95\%$ confidence level $r \lesssim 0.05$). Therefore, by fixing one (or two) parameters and taking a sampling of well-distributed points in the resultant subspace gives us the conditional likelihood that the model can match observations. 
 
Maximizing our ignorance, we choose a uniform distribution for all the parameters, and in the interest of naturalness, we further restrict the parameters in the range $[-1,1]$. To begin with, we examine the viability of the theory without \emph{a priori} constraining any parameters by random sampling $10^6$ points, and present the results in Table~\ref{tab:conditions}. Looking at the table, it appears that the no-ghost condition is the most restrictive of the three conditions, and relaxing it contributes the most significantly to broadening the viable regions of the theory. This is reasonable since inflation occurs in a relatively small region inside the different domains demarcated by the poles, and it is an unnecessary restriction to assume that there must be no ghosts in a domain that is after all not visited by the field. We therefore find that  requiring that there are no ghosts near the poles (i.e. imposing the more relaxed no-ghost condition) leads to a rather robust model with a much higher likelihood of viability (for a uniform distribution on the parameters). 

\begin{figure}[t]
\centering
\includegraphics[scale=0.5]{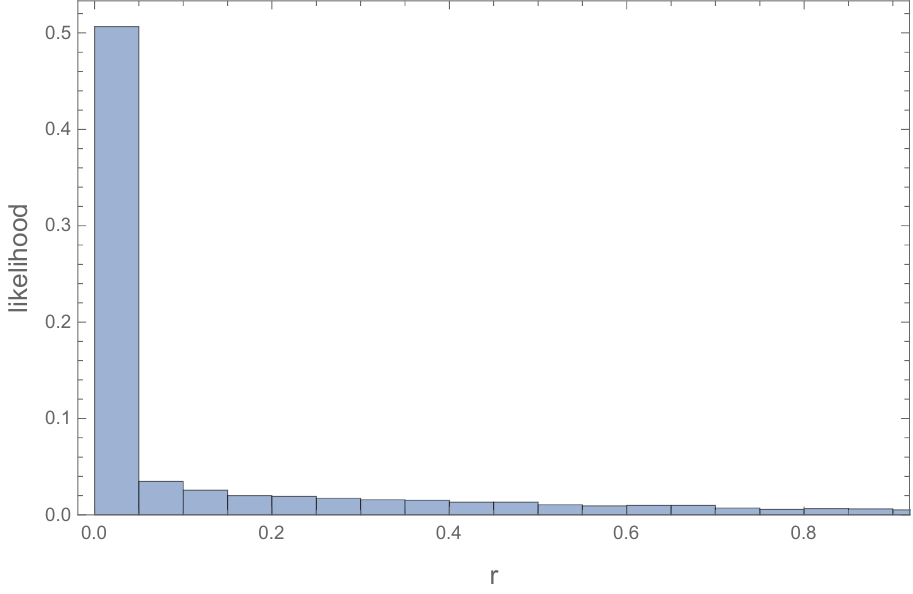}
\includegraphics[scale=0.5]{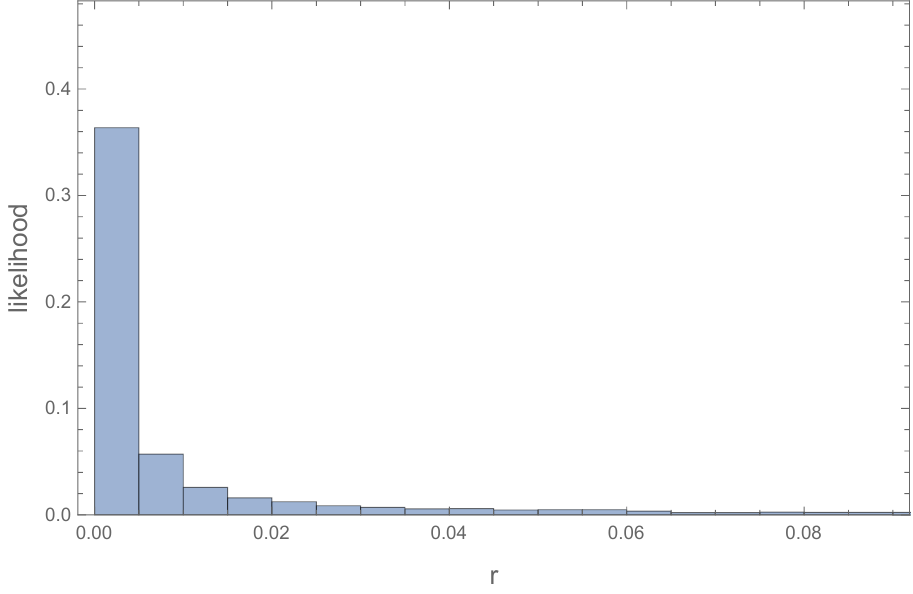}
\caption{Likelihood of obtaining different values of $r$ with a uniform sampling of parameters, given that the second-order pole and the near no-ghost conditions are satisfied. The left plot shows a range of $[0,1]$ and the right plot the more relevant range of $[0,0.1]$.}\label{fig5}
\end{figure}

Focusing on the more relaxed no-ghost condition, we can chart the predictions for the tensor-to-scalar ratio $r$ of the points in the parameter space. We find that about $75.6\%$ of the points that satisfy both the second-order pole and the near no-ghost condition predict a tensor-to-scalar ratio $r<1$. For the rest of the points, this indicates that the residue of the pole is so large that the slow-roll condition is violated very close to the pole). Still, we find that most values are under $0.1$: this occurs for about $54.1\%$ of points, and in fact the distribution of tensor-to-scalar predictions shows a clear preference for low values of $r$. We present the distribution of tensor-to-scalar predictions up until $r=1$ in Figure~\ref{fig5}.

\begin{table}
\centering
\begin{tabular}{ l@{\hskip 0.5cm}lllr }
condition & likelihood   \\
\hline
second-order pole & 0.8254      \\
second-order pole, no ghost & 0.001184   \\
second-order pole, no ghost near pole &  0.8213  \\
%\\
%second-order pole, no ghost, $r<0.05$& 0.001184
%second order pole, no ghost near pole, $r<0.05$ & 0.8228
\end{tabular}
\caption{ Likelihood of satisfying different conditions for a uniformly random selection of parameters within $[-1,1]$. The no-ghost condition is checked for at least one of the second-order poles.}
\label{tab:conditions}
\end{table}

\begin{figure}[!ht]
     \centering
    \begin{tabular}{ccc}
        \includegraphics[scale=0.4]{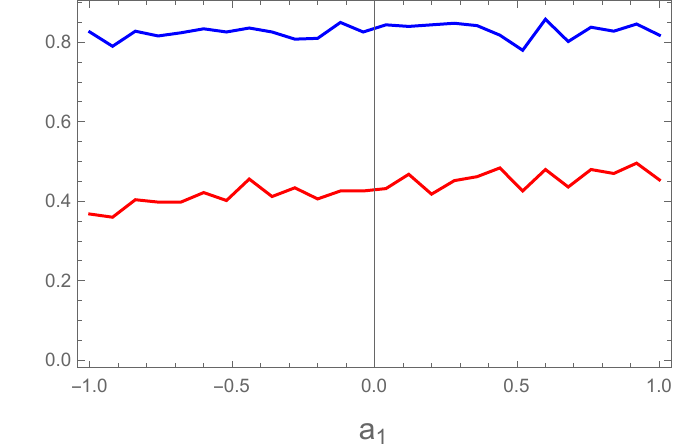} &
        \includegraphics[scale=0.4]{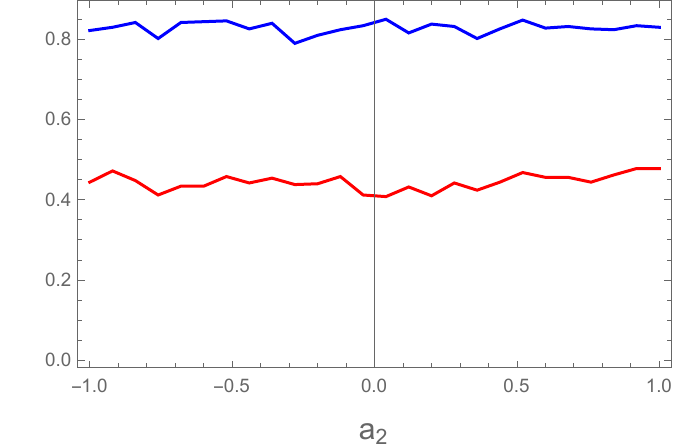} &
        \includegraphics[scale=0.4]{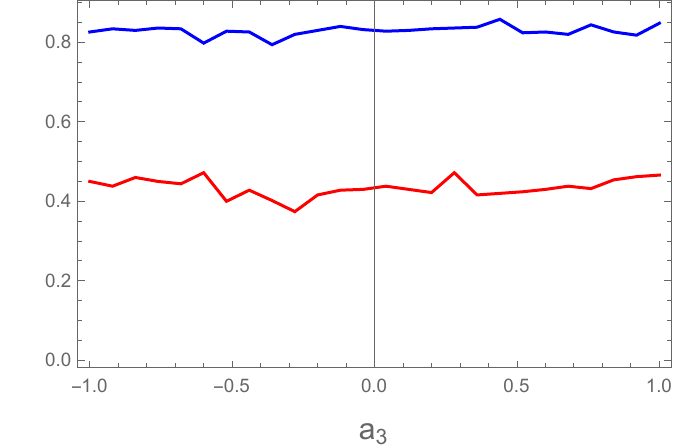} \\[1ex]

        \multicolumn{3}{c}{
            \includegraphics[scale=0.4]{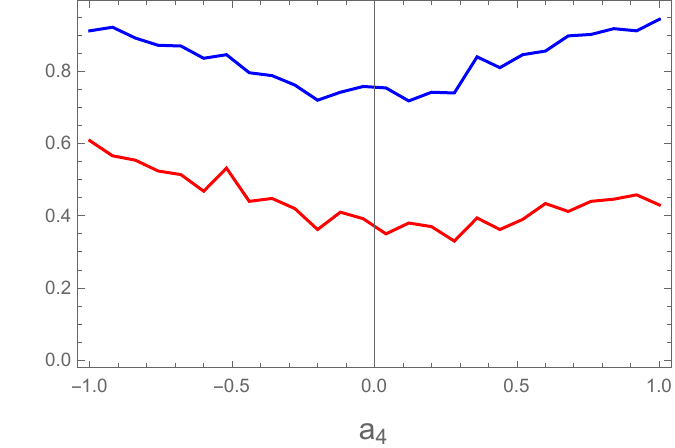} \hspace{1em}
            \includegraphics[scale=0.4]{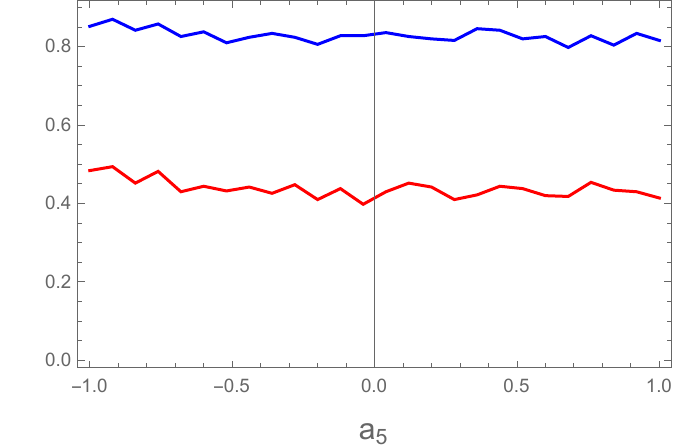}
        } \\[1ex]

        \includegraphics[scale=0.4]{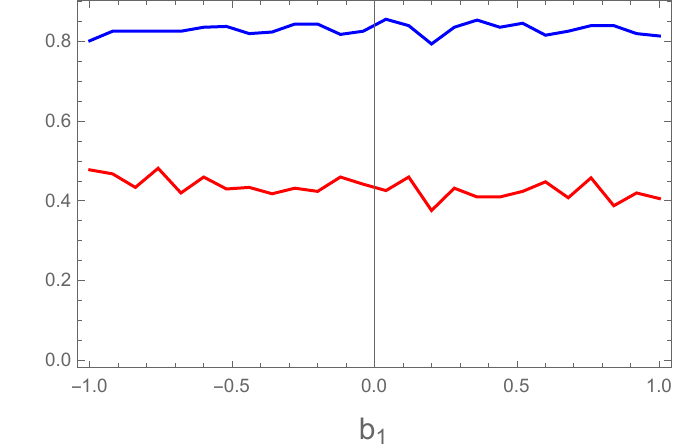} &
        \includegraphics[scale=0.4]{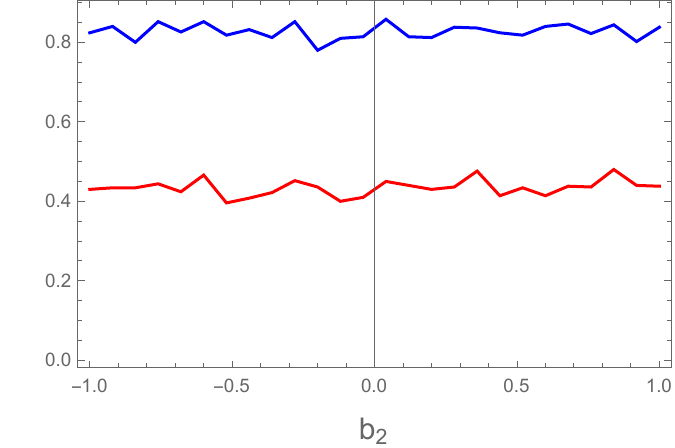} &
        \includegraphics[scale=0.35]{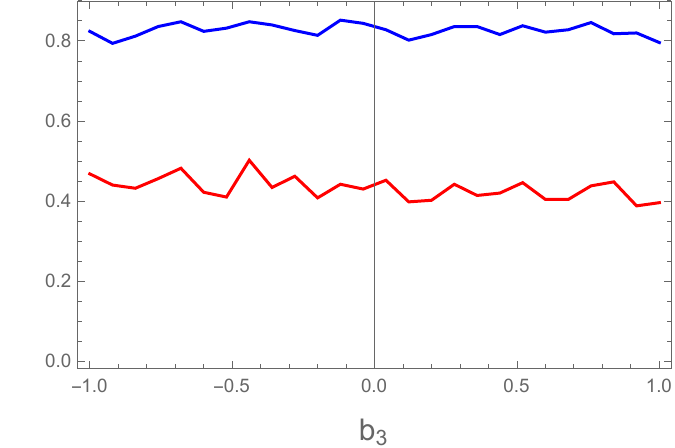} \\[1ex]

        \includegraphics[scale=0.4]{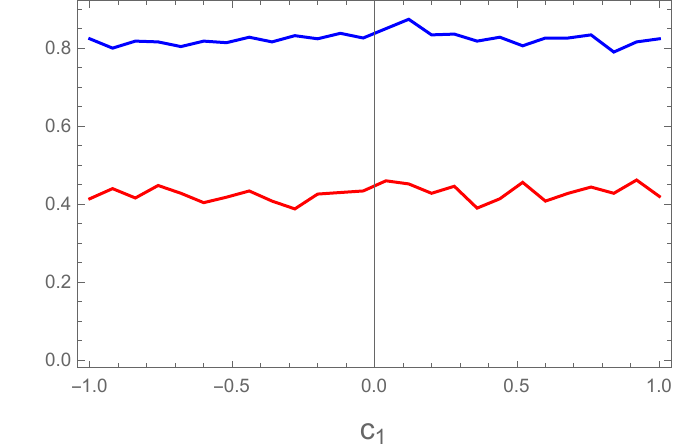} &
        \includegraphics[scale=0.4]{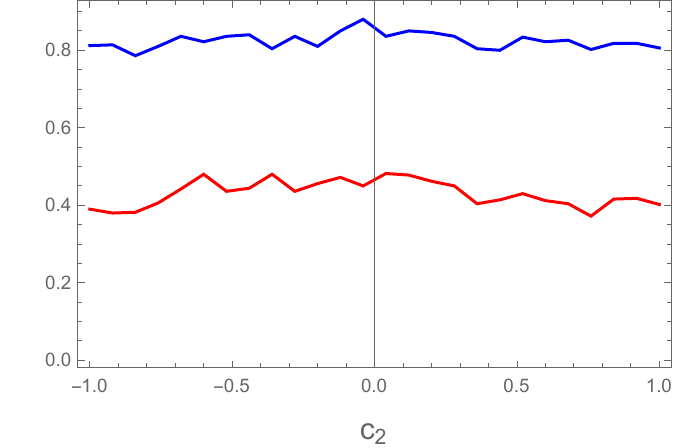} &
        \includegraphics[scale=0.4]{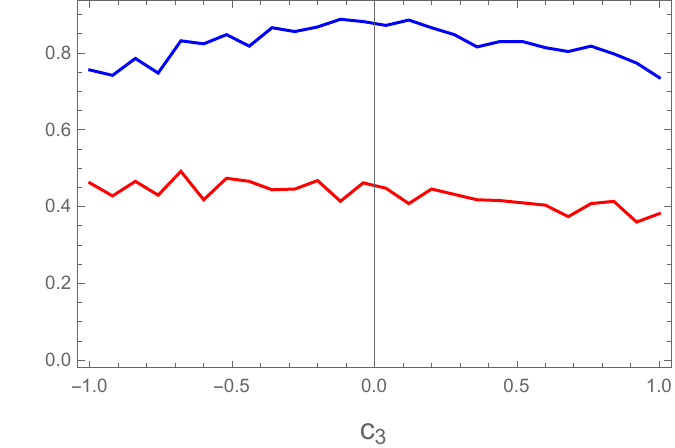} \\
    \end{tabular}
     
    \caption{Marginalized likelihood of having both a second-order pole and having no ghosts near the poles (blue line) and also predicting $r\lesssim 0.05$ (red line) over the rest of the parameters (uniformly chosen in the same range of [-1,1]).}
    \label{fig:nearmarginalizedconditions}
\end{figure}

So far, we have presented the likelihood for satisfying the various conditions under the assumption that all 11 parameters are distributed according to their full (uniform) joint distribution. We can further examine the likelihood functions that arise all but one of the parameters are marginalized. We first present the marginalized likelihood functions for satisfying the second-order pole condition together with the near no-ghost condition, as well as the observational viability condition at $N_\star = 60$ with a limit of $r < 0.05$, shown in Figure~\ref{fig:nearmarginalizedconditions} for $10^5$ sample points. We observe that the viability of the theory is mostly insensitive to the choice of parameters, save for $a_4$, $c_2$, and $c_3$, which do not demonstrate a flat function. This indicates that the general model is fairly robust as long as we are willing to relax the no-ghost condition as described above, which intuitively agrees with the fact that roughly half of the sample points (that satisfy the rest of the conditions) also satisfy the observational constraint as shown in Fig.~\ref{fig5}.
  
\begin{figure}[!ht]
    \centering
    \begin{tabular}{ccc}
        \includegraphics[scale=0.5]{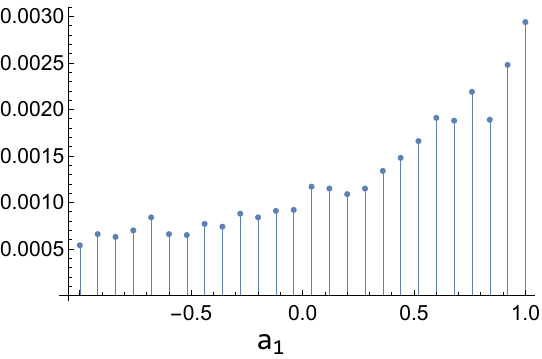} &
        \includegraphics[scale=0.5]{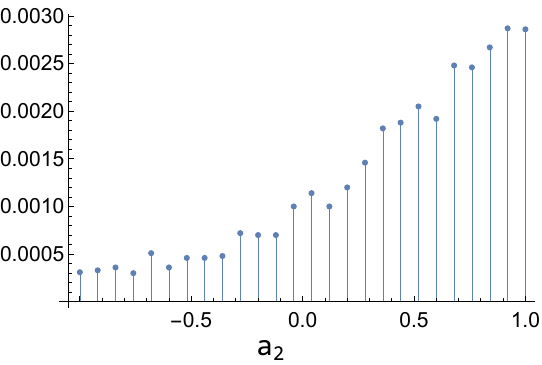} &
        \includegraphics[scale=0.5]{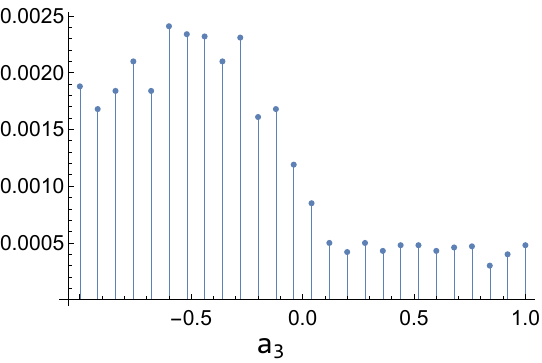} \\[1ex]

        \multicolumn{3}{c}{
            \includegraphics[scale=0.5]{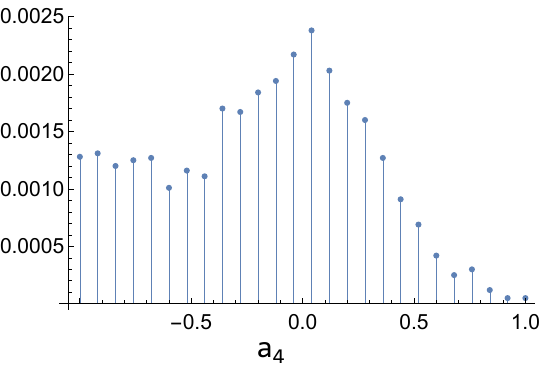} \hspace{1em}
            \includegraphics[scale=0.5]{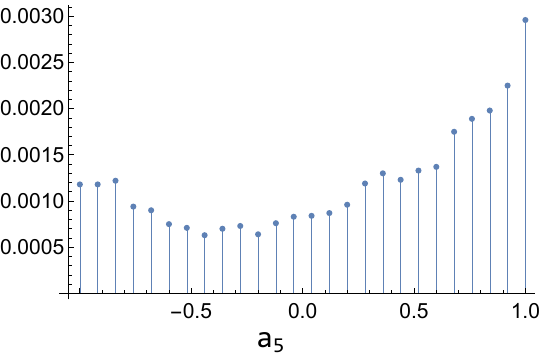}
        } \\[1ex]

        \includegraphics[scale=0.5]{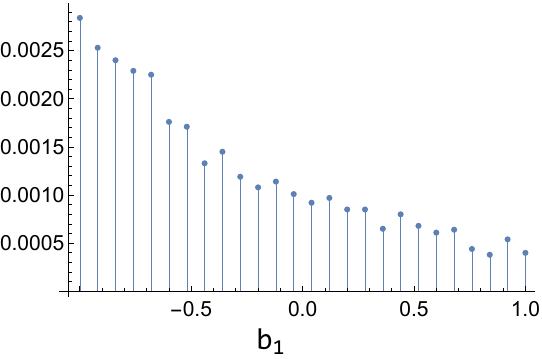} &
        \includegraphics[scale=0.5]{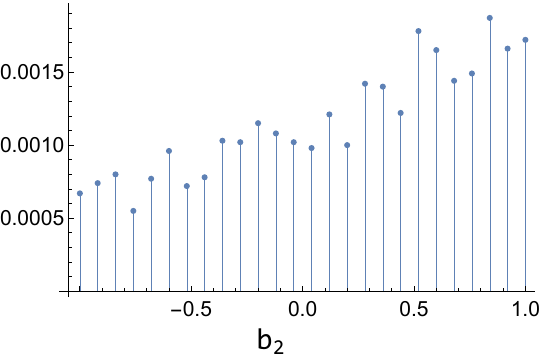} &
        \includegraphics[scale=0.5]{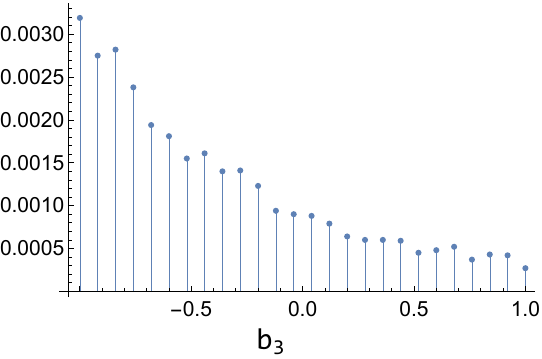} \\[1ex]

        \includegraphics[scale=0.5]{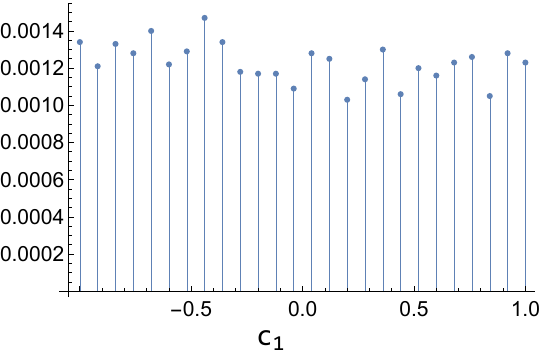} &
        \includegraphics[scale=0.5]{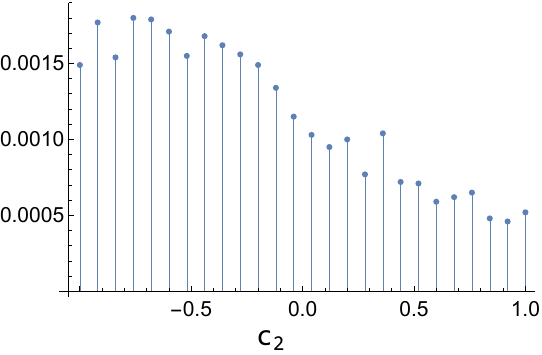} &
        \includegraphics[scale=0.5]{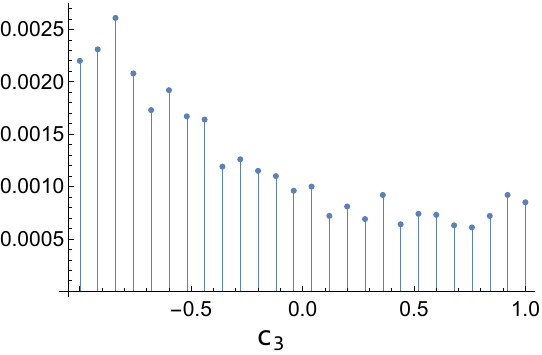} \\
    \end{tabular}
    \caption{Marginalized likelihood of satisfying $r\lesssim 0.05$ over the rest of the parameters uniformly chosen in the same range of [-1,1], where we have imposed the strict no-ghost condition.}
    \label{fig:nearmarginalizedconditionsstrict}
\end{figure}

\begin{figure}[!ht]
\centering
\includegraphics[scale=0.13]{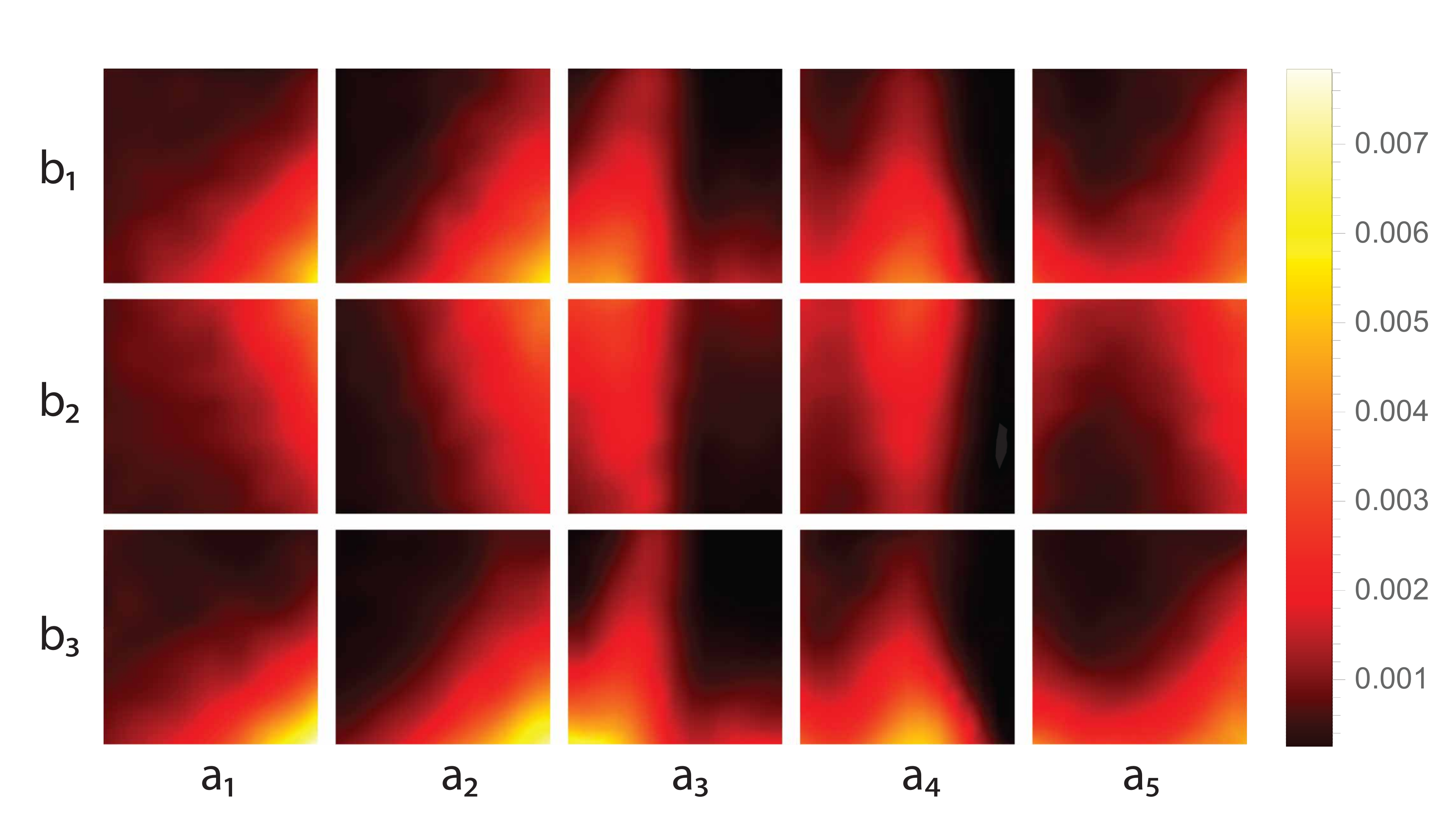}
\includegraphics[scale=0.13]{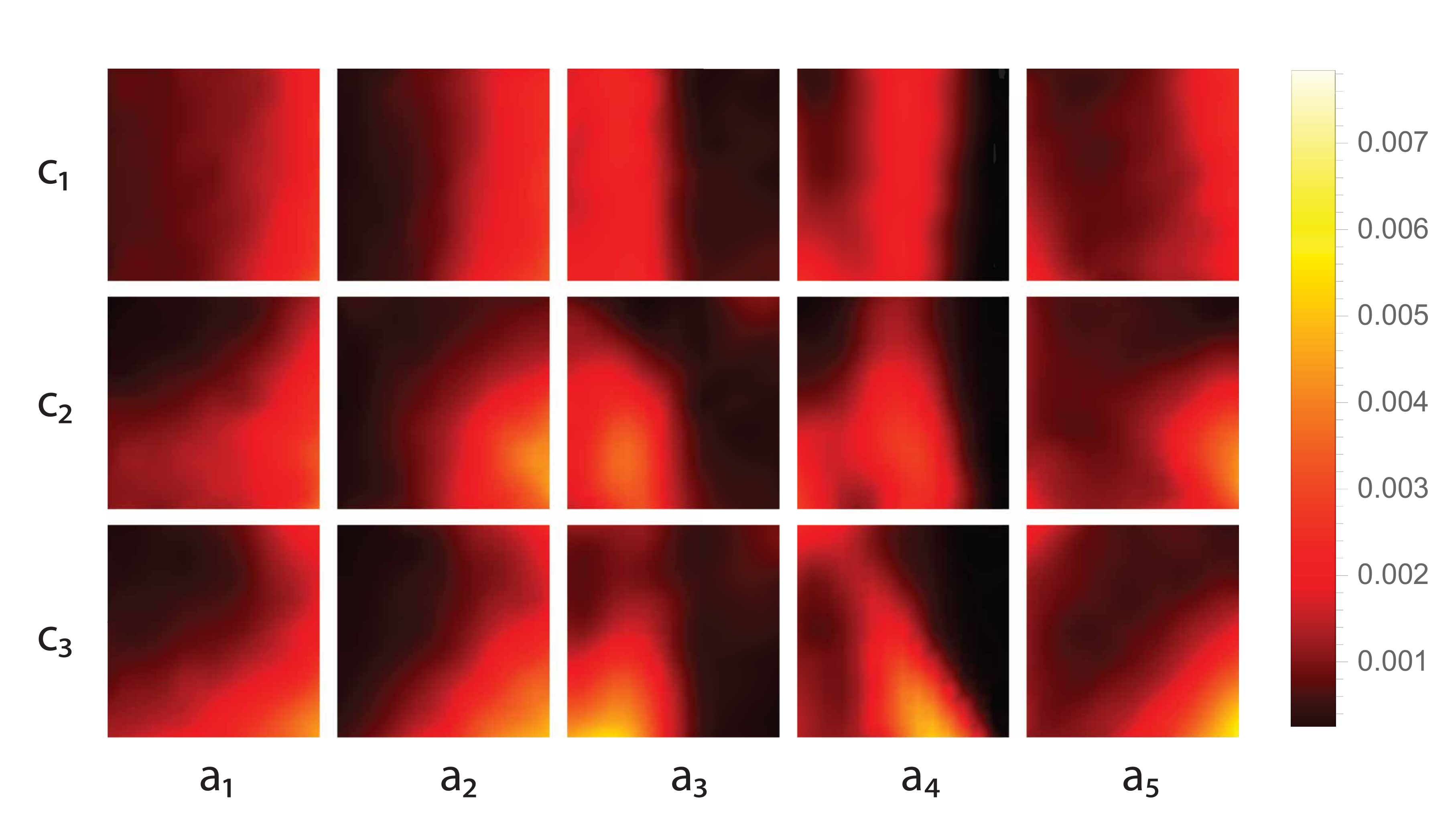}
\includegraphics[scale=0.13]{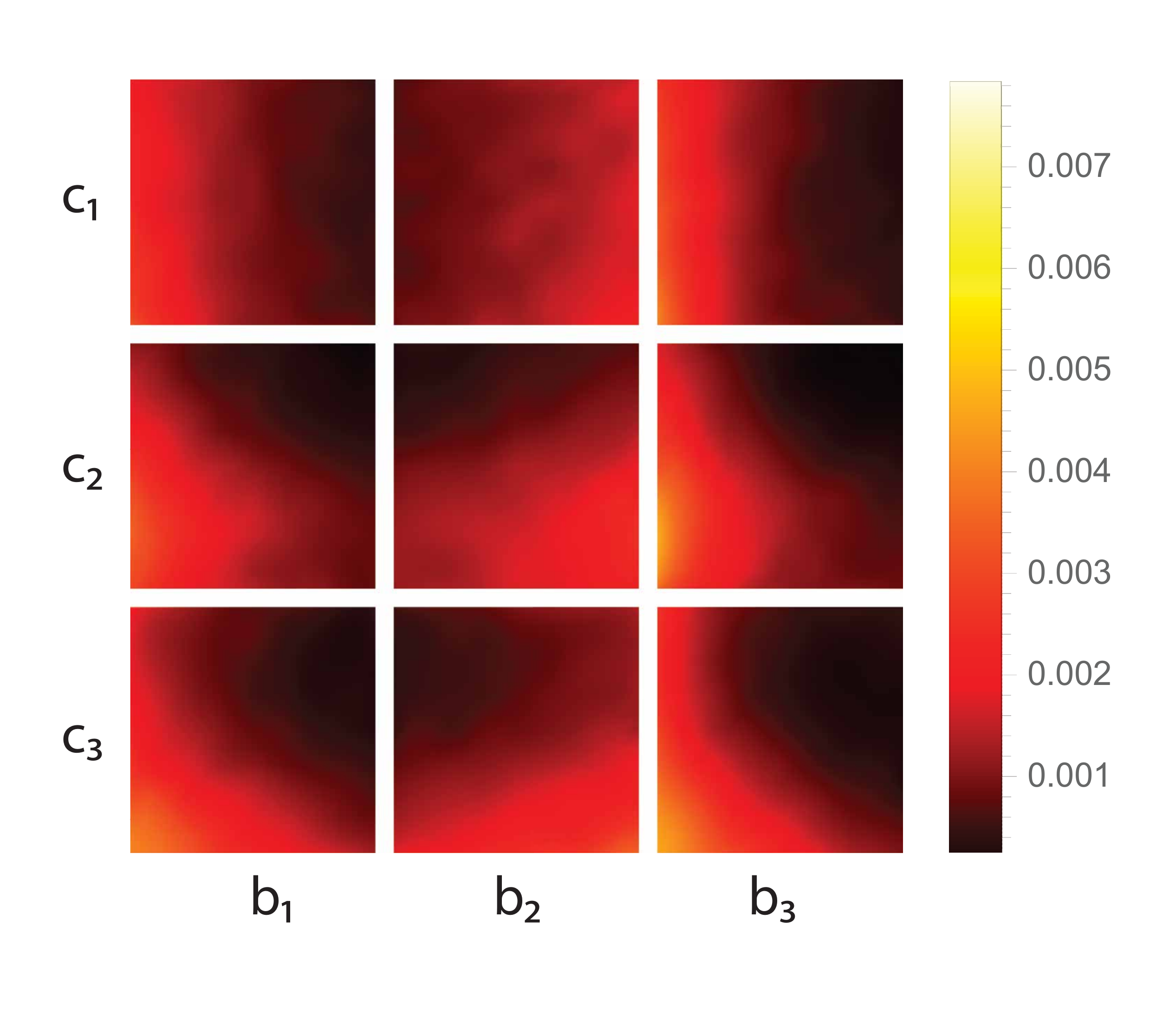}
\includegraphics[scale=0.13]{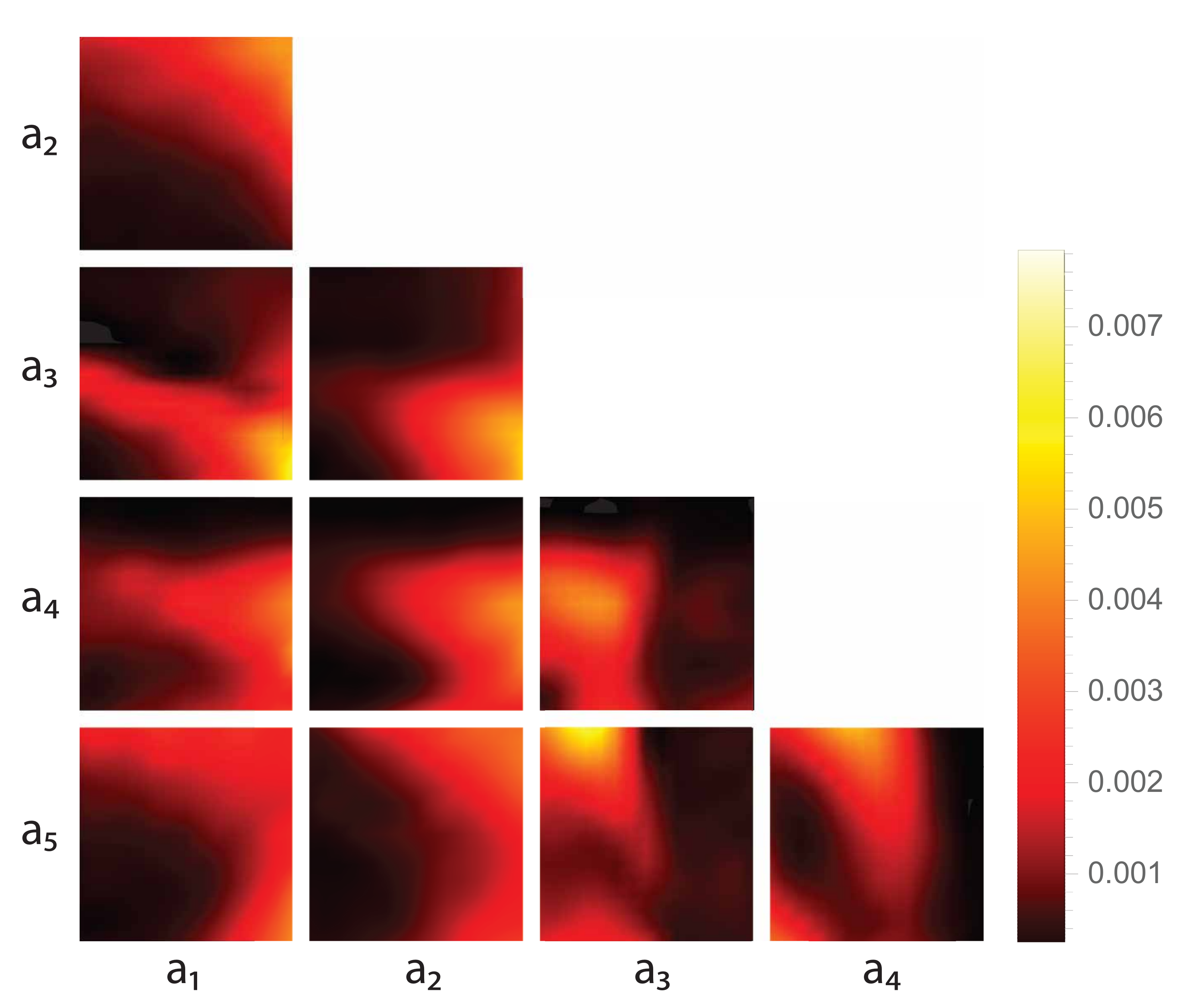}
\includegraphics[scale=0.13]{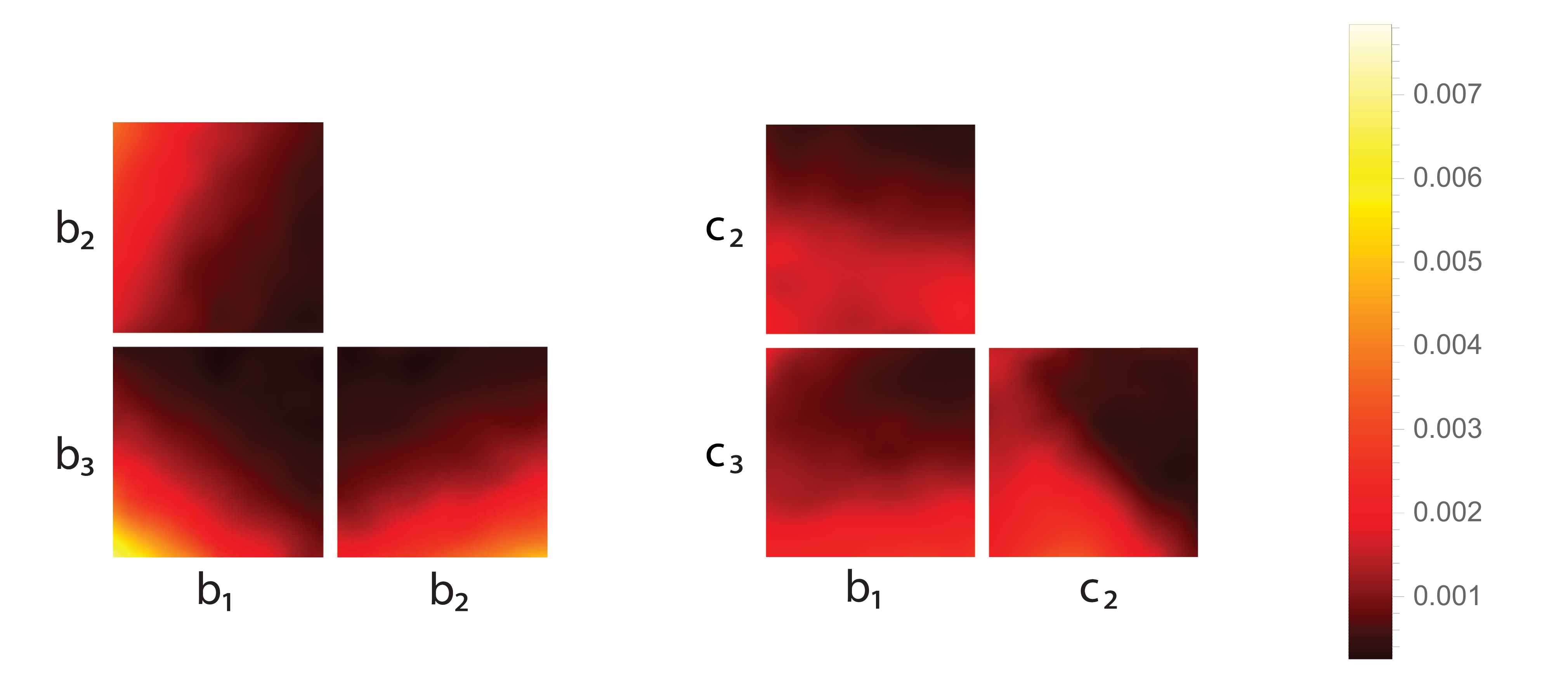}
\caption{  
Likelihood of satisfying observational constraints (with a strict no-ghost condition) for fixed pairs of parameters $(a_i,b_j)$, with the rest of the parameters uniformly distributed over $[-1,1]$. For each subgraph, the $x$-axis runs over the parameter shown in the column and the $y$-axis runs over the parameter shown in the row. Lighter areas correspond to higher likelihood.}\label{fig:joint}
\end{figure}

While the relaxed no-ghost condition is perhaps more physically relevant, it is interesting to see the marginalized likelihood of satisfying $r\lesssim 0.05$ under the stricter no-ghost condition. We present the likelihood functions in Fig.~\ref{fig:nearmarginalizedconditionsstrict} (again sampled at $10^5$ points), where we immediately note that the probabilities are much smaller, which again agrees with our observation that most choices of parameters will feature ghosts (and so only a very small proportion of those can lead to viable inflation), but most of these ghosts do not appear close to the poles (where the kinetic term remains positive). Unlike the  relaxed no-ghost condition, there are values that are clearly favored, and so for the strict no-ghost condition, we also plot the joint likelihood function in terms of two parameters, as seen in Fig.~\ref{fig:joint}. We observe that similarly to the single-parameter likelihood distributions, there are certain areas that are excluded if we rule out ghosts entirely.

\subsection{Special cases}

Let us gather here some special cases where the coefficients of the quadratic terms are those of the teleparallel equivalents of GR. We will examine the following subcases: 
\begin{itemize}
\item {\bf Metric teleparallelism} \cite{DeAndrade:2000sf,BeltranJimenez:2018vdo}: 
\begin{align}
b_{1}=1 \;, b_{2}=-2\;, b_{3}=-4\; , \;a_{i}=0=c_{i}
\end{align}

\item {\bf Symmetric teleparallelism} \cite{Nester:1998mp,BeltranJimenez:2018vdo}: 
\begin{align}
a_{1}=\frac{1}{4} \;, a_{2}=-\frac{1}{2}\;, a_{3}=-\frac{1}{4}\; , a_{4}=0 \;, a_{5}=\frac{1}{2}\;, \;b_{i}=0=c_{i}
\end{align}

\item {\bf General teleparallelism \cite{BeltranJimenez:2019odq}:}

\begin{align}
b_{1}=1 \;, b_{2}=-2\;, b_{3}=-4\; , \;a_{1}=\frac{1}{4} \;, a_{2}=-\frac{1}{2}\;, a_{3}=-\frac{1}{4}\; , a_{4}=0 \;, a_{5}=\frac{1}{2}\;, \; c_{1}=-c_{2}=c_{3}=2
\end{align}
\end{itemize}
We note that we could introduce an overall $-1$ sign in these parameter sets, although we find that the final results are unaffected due to symmetry. We summarize the results in Table~\ref{tab:specialcases}, where we also show the predicted tensor-to-scalar ratio for $N=60$ as well as $N=70$. In particular, the latter pushes the prediction of $n_s$ closer to the ACT contour as well as drive $r$ down, although it corresponds to a lower reheating temperature through the relation \cite{Cook:2015vqa, Becker:2023tvd}
\begin{align}
T_{\rm rh} = \left[ 
\left( \frac{43}{11\, g_{\ast } } \right)^{\!1/3} 
\frac{a_0 T_0}{k_\star} H_\star e^{-N_\star} 
\left( \frac{45 V_{\rm end}}{\pi^2 g_\ast } \right)^{-\tfrac{1}{3(1+w_{\rm eff})}} 
\right]^{\tfrac{3(1+w_{\rm eff})}{3 w_{\rm eff} - 1}},
\end{align}
where the effective number of relativistic degrees of freedom is mostly insensitive for temperatures above $\sim 200 \ {\rm GeV}$. Still, a careful treatment for which the degrees of freedom depend on the temperature, and for $w_{\rm eff} >0$ (i.e. departing from pure perturbative reheating), the maximum reheating temperature emerges as $T_{\rm reh} \lesssim 10^{10} \ {\rm GeV}$, which corresponds to a range for the number of e-folds $58 \lesssim N_\star \lesssim 69 $ \cite{Drees:2025ngb}.  

\begin{table}[!hb]
\centering
\begin{tabular}{ l@{\hskip 0.5cm}lllll }
condition & pole & no ghost & no near ghost & $r (N_\star=60)$  & $r (N_\star=70)$ \\
\hline
metric TP & no & - & - & - & - \\
symmetric TP & yes & no & yes & 0.0168 & 0.0123\\
general TP & yes & no & yes &  $8.24 \times 10^{-4}$ & $6.05 \times 10^{-4}$
\end{tabular}
\caption{Different special cases for the general model and whether they satisfy the different conditions along with predicted values for $r$.}
\label{tab:specialcases}
\end{table}

\section{Conclusion}
\label{conclusion}

In this work, we examined the most general class of extended metric-affine F(R) models featuring quadratic torsion and non-metricity invariants, focusing on their realization of pole inflation that follows as a result of their scalar degree of freedom which acquires a kinetic term with a prefactor that blows up at certain field values. Scanning over the parameter space for a simplified model, we identified the subsets that support genuine second-order poles (therefore deviating from power-law inflation). We then imposed successively stronger requirements, first the absence of ghosts, and then observational viability of the tensor-to-scalar ratio. We found that a sizable portion of the parameter space survives all three filters, showing that attractor inflation from such general models remains consistent with constraints on $r$ as well as the recent constraints on $n_s$ from ACT if a lower reheating temperature is allowed. We also found that the GR equivalents of symmetric and general teleparallelism endowed with an arbitrary $F(R)$ (such that they deviate from GR) can be observationally viable, although metric teleparallelism cannot.

We found that the biggest stopgap in the viability of the different models captured by the parameter space is the existence of ghosts: a strict condition that excludes all ghosts by demanding that the kinetic term remains positive everywhere within the domain where the scalar field is constrained. However, allowing for ghosts in more distant regions away from the pole drastically increases the proportion of viable theories, since the presence of ghosts in a domain which the field will not traverse during inflation no longer unnecessarily excludes models which give rise to viable predictions. With this reasonable relaxation of the no-ghost condition, the extended metric-affine $F(R)$ models display robustness both due to the lack of sensitivity on the specific form of $F(R)$ (which sets the form of the potential that pole inflation is insensitive to) as well as due to the majority of models as specified by the coefficients of the torsion/non-metricity invariants. 
 
Finally, the dynamics of torsion and non-metricity throughout the inflationary regime emerge as byproduct of our analysis. Their evolution can be studied both within the slow-roll approximation and numerically. Their behavior appears to be tied to the pole structure that emerges from this extended model. Exploring their behavior during inflation as well as in the post-inflationary period is an interesting direction to study, as examples of non-Riemannian quantities in early-universe cosmology with a connection to observable quantities.

\section*{Acknowledgements}

DI acknowledges  the support of Istituto Nazionale di Fisica Nucleare (INFN), Sezioni di Napoli, Iniziative Specifiche QGSKY. SK was funded by the Estonian Research Council Mobilitas 3.0 incoming postdoctoral grant MOB3JD1233 ``Inflationary Nonminimal Models: An Investigative Exploration''.

\appendix
\section{Explicit expression for all constants}
\label{expl}

We present the definitions of all constants sequentially, such that each definition only includes constants defined in the previous subsections. Thus, every constant can be written in terms of the eleven constants $a_i, 
 b_i$, and $c_i$.

\subsection{$a_{ij}, b_{ij}, c_{ij}$}

\begin{align*}
a_{11} &= 4 a_1 - \frac{c_1}{2} + 4 n a_3 + 2 a_5 + \frac{1 - n}{2} c_2, \\
a_{12} &= 4 a_2 + \frac{c_1}{2} + 2 n a_5 + 4 a_4 + \frac{1 - n}{2} c_3, \\
a_{13} &= -2 b_1 + b_2 + 2 c_1 + 2 n c_2 + 2 c_3 + (1 - n) b_3, \\
a_{21} &= -\frac{(n - 1)}{2} f'(\Phi) + 2 a_2 + \frac{c_1}{2} + 4 a_3 + (n + 1) a_5 + \frac{n - 1}{2} c_2, \\
a_{22} &= (n - 1) f'(\Phi) + 4 a_1 + 2 a_2 - \frac{c_1}{2} + 2 (n + 1) a_4 + 2 a_5 + \frac{n - 1}{2} c_3\\
a_{23} &= 2 (2 - n) f'(\Phi) + 2 b_1 - b_2 - c_1 + 2 c_2 + (n + 1) c_3 + (n - 1) b_3, \\
a_{31} &= \frac{(n - 3)}{2} f'(\Phi) + 2 a_2 + 4 a_3 + (n + 1) a_5, \\
a_{32} &= f'(\Phi) + 2 (2 a_1 + a_2 + a_5 + (n + 1) a_4), \\
a_{33} &= 2 (n - 2) f'(\Phi) - c_1 + 2 c_2 + (n + 1) c_3, \\
b_{21} &= -\frac{n - 1}{2}, \\
c_{21} &= 2 a_2 + \frac{c_1}{2} + 4 a_3 + (n + 1) a_5 + \frac{n - 1}{2} c_2, \\
b_{22} &= n - 1, \\
c_{22} &= 4 a_1 + 2 a_2 - \frac{c_1}{2} + 2 (n + 1) a_4 + 2 a_5 + \frac{n - 1}{2} c_3, \\
b_{23} &= 2 (2 - n), \\
c_{23} &= 2 b_1 - b_2 - c_1 + 2 c_2 + (n + 1) c_3 + (n - 1) b_3, \\
b_{31} &= \frac{n - 3}{2}, \\
c_{31} &= 2 a_2 + 4 a_3 + (n + 1) a_5, \\
b_{32} &= 1, \\
c_{32} &= 2 (2 a_1 + a_2 + a_5 + (n + 1) a_4), \\
b_{33} &= 2 (n - 2), \\
c_{33} &= -c_1 + 2 c_2 + (n + 1) c_3.
\end{align*}

\subsection{$\beta_{ij}$}

\begin{align*}
\beta_{11} &= 2 a_2 + 4 a_4 - \frac{b_1}{2} + \frac{b_2}{2} + \frac{b_3}{4} 
+ 4 (a_1 + \frac{b_1}{2} - c_1) 
+ 4 (a_1 + a_2 - \frac{b_2}{2} - c_1) 
+ 2 c_1 + 16 (4 a_3 + \frac{b_3}{4} - c_2) 
\\
&\hphantom{=} + 4 (a_5 - \frac{c_3}{2}) 
+ 4 (a_5 - \frac{b_3}{2} + c_2 - \frac{c_3}{2}) 
+ \frac{3 c_3}{2}, \\
\beta_{12} &= 17 + 2 a_4 + 8 (4 a_3 + \frac{b_3}{4} - c_2) 
+ 5 (a_5 - \frac{c_3}{2}) 
+ 17 (a_5 - \frac{b_3}{2} + c_2 - \frac{c_3}{2}) 
+ 5 (2 a_4 + \frac{c_3}{2}) 
+ 2 (a_4 + \frac{b_3}{4} + c_3), \\
\beta_{13} &= 17 + 8 a_4 + 8 (4 a_3 + \frac{b_3}{4} - c_2) 
+ 17 (a_5 - \frac{c_3}{2}) 
+ 5 (a_5 - \frac{b_3}{2} + c_2 - \frac{c_3}{2}) 
+ 5 (2 a_4 + \frac{c_3}{2}) 
+ 2 (a_4 + \frac{b_3}{4} + c_3),
\end{align*}

\begin{align*}
\beta_{21} &= 0, \\
\beta_{22} &= 2 a_1 + 3 a_2 + 4 a_3 + 2 a_4 + a_5 + \frac{b_3}{2} - c_2 
+ 4 (a_5 - \frac{b_3}{2} + c_2 - \frac{c_3}{2}) 
+ 4 (2 a_4 + \frac{c_3}{2}) 
+ \frac{c_3}{2}, \\
\beta_{23} &= 17 + 8 a_4 + 2 (4 a_3 + \frac{b_3}{4} - c_2) 
+ 5 (a_5 - \frac{c_3}{2}) 
+ 5 (a_5 - \frac{b_3}{2} + c_2 - \frac{c_3}{2}) 
+ 17 (2 a_4 + \frac{c_3}{2}) 
+ 2 (a_4 + \frac{b_3}{4} + c_3), \\
\beta_{31} &= 0, \\
\beta_{32} &= 0, \\
\beta_{33} &= 2 a_1 + 3 a_2 + 4 a_3 + 17 a_4 + a_5 
+ 4 (a_5 - \frac{c_3}{2}) 
+ 4 (2 a_4 + \frac{c_3}{2}) 
+ \frac{c_3}{2}.
\end{align*}

\subsection{$A_0, B_0, C_0$}

\begin{align*}
A_0 &= a_{13} b_{22} b_{31} + a_{12} b_{23} b_{31} + a_{13} b_{21} b_{32} 
- a_{11} b_{23} b_{32} - a_{12} b_{21} b_{33} + a_{11} b_{22} b_{33}, \\
B_0 &= a_{13} b_{32} c_{21} - a_{12} b_{33} c_{21} - a_{13} b_{31} c_{22} 
+ a_{11} b_{33} c_{22} + a_{12} b_{31} c_{23} - a_{11} b_{32} c_{23} \\
&\quad - a_{13} b_{22} c_{31} + a_{12} b_{23} c_{31} + a_{13} b_{21} c_{32} 
- a_{11} b_{23} c_{32} - a_{12} b_{21} c_{33} + a_{11} b_{22} c_{33}, \\
C_0 &= -a_{13} c_{22} c_{31} + a_{12} c_{23} c_{31} + a_{13} c_{21} c_{32} 
- a_{11} c_{23} c_{32} - a_{12} c_{21} c_{33} + a_{11} c_{22} c_{33}.
\end{align*}

\subsection{$\widetilde A_{ij}, \widetilde B_{ij}$}

\begin{align*}
\widetilde{A}_{12} &= a_{13} b_{32} - a_{12} b_{33}, & 
\widetilde{B}_{12} &= a_{13} c_{32} - a_{12} c_{33}, \\
\widetilde{A}_{13} &= -a_{13} b_{22} + a_{12} b_{23}, & 
\widetilde{B}_{13} &= -a_{13} c_{22} + a_{12} c_{23}, \\
\widetilde{A}_{22} &= -a_{13} b_{31} + a_{11} b_{33}, & 
\widetilde{B}_{22} &= -a_{13} c_{31} + a_{11} c_{33}, \\
\widetilde{A}_{23} &= a_{13} b_{21} - a_{11} b_{23}, & 
\widetilde{B}_{23} &= a_{13} c_{21} - a_{11} c_{23}, \\
\widetilde{A}_{32} &= a_{12} b_{31} - a_{11} b_{32}, & 
\widetilde{B}_{32} &= a_{12} c_{31} - a_{11} c_{32}, \\
\widetilde{A}_{33} &= -a_{12} b_{21} + a_{11} b_{22}, & 
\widetilde{B}_{33} &= -a_{12} c_{21} + a_{11} c_{22}.
\end{align*}

\subsection{$A_i, B_i$}

\begin{align*}
A_1 &= (n - 1) (\widetilde{A}_{13} - \widetilde{A}_{12}), & 
B_1 &= (n - 1) (\widetilde{B}_{13} - \widetilde{A}_{12}), \\
A_2 &= (n - 1) (\widetilde{A}_{23} - \widetilde{A}_{22}), & 
B_2 &= (n - 1) (\widetilde{B}_{23} - \widetilde{B}_{22}), \\
A_3 &= (n - 1) (\widetilde{A}_{33} - \widetilde{A}_{32}), & 
B_3 &= (n - 1) (\widetilde{B}_{33} - \widetilde{B}_{32}), \\
A_4 &= \frac{A_1}{4}, & 
B_4 &= \frac{B_1}{4}, \\
A_5 &= \frac{A_1}{2} + 2 A_3, & 
B_5 &= \frac{B_1}{2} + 2 B_3, \\
A_6 &= A_2 - \frac{A_1}{2} - 2 A_3, & 
B_6 &= B_2 - \frac{B_1}{2} - 2 B_3.
\end{align*}

\subsection{$C_i, D_i$}

\begin{align*}
C_1 &= \frac{(n + 1) A_4 - A_5 - A_6}{(n - 1)(n + 2)}, & 
D_1 &= \frac{(n + 1) B_4 - B_5 - B_6}{(n - 1)(n + 2)}, \\
C_2 &= \frac{-A_4 + (n + 1) A_5 - A_6}{(n - 1)(n + 2)}, & 
D_2 &= \frac{-B_4 + (n + 1) B_5 - B_6}{(n - 1)(n + 2)}, \\
C_3 &= \frac{-A_4 - A_5 + (n + 1) A_6}{(n - 1)(n + 2)}, & 
D_3 &= \frac{-B_4 - B_5 + (n + 1) B_6}{(n - 1)(n + 2)}.
\end{align*}

\subsection{$k_i$}
\begin{align*}
k_1(\Phi) &=   (C_2 \Phi + D_2)^2 + (C_3 \Phi + D_3)^2 + 
n (C_2 \Phi + D_2)(C_3 \Phi + D_3) , \\
k_2(\Phi) &=   (C_2 - C_3) \Phi + (D_2 - D_3)  , \\
k_3(\Phi) &=   (C_1 \Phi + D_1) \beta_{11} (C_1 \Phi + D_1) + 
(C_2 \Phi + D_2) \beta_{21} (C_1 \Phi + D_1) + 
(C_3 \Phi + D_3) \beta_{31} (C_1 \Phi + D_1) \\
& +(C_1 \Phi + D_1) \beta_{12} (C_2 \Phi + D_2) + 
(C_2 \Phi + D_2) \beta_{22} (C_2 \Phi + D_2) + 
(C_3 \Phi + D_3) \beta_{32} (C_2 \Phi + D_2)  \\
&+ (C_1 \Phi + D_1) \beta_{13} (C_3 \Phi + D_3) + 
(C_2 \Phi + D_2) \beta_{23} (C_3 \Phi + D_3) + 
(C_3 \Phi + D_3) \beta_{33} (C_3 \Phi + D_3) .
\end{align*}

\subsection{$\mu_i$}\label{mus}

\begin{align*}
\mu_1 &= 
2 [ 3 C_0 D_2 - 3 C_0 D_3 + D_1^2 \beta_{11} + D_1 D_2 \beta_{12} + D_1 D_3 \beta_{13} + D_1 D_2 \beta_{21} + D_2^2 \beta_{22} 
\\
&\hphantom{=}+ D_2 D_3 \beta_{23} + D_1 D_3 \beta_{31} + D_2 D_3 \beta_{32} + D_3^2 \beta_{33} ]
 \\
\mu_2 &= 
2 [ 3 C_0 (C_2 - C_3) + 3 B_0 D_2 + 3 D_2^2 - 3 B_0 D_3 + 12 D_2 D_3 + 3 D_3^2 + 
2 C_1 D_1 \beta_{11} + C_2 D_1 \beta_{12} + C_1 D_2 \beta_{12} 
\\
&\hphantom{=}+ 
C_3 D_1 \beta_{13} + C_1 D_3 \beta_{13} + C_2 D_1 \beta_{21} +  C_1 D_2 \beta_{21} + 2 C_2 D_2 \beta_{22} + C_3 D_2 \beta_{23} + 
C_2 D_3 \beta_{23} + C_3 D_1 \beta_{31} + C_1 D_3 \beta_{31} 
\\
&\hphantom{=}+
C_3 D_2 \beta_{32} + C_2 D_3 \beta_{32} + 2 C_3 D_3 \beta_{33} ]
\\
\mu_3 &= 
2 [ 3 B_0 (C_2 - C_3) + 3 A_0 D_2 + 6 C_2 D_2 + 12 C_3 D_2 - 3 A_0 D_3 + 12 C_2 D_3 + 6 C_3 D_3 + C_1^2 \beta_{11} + C_1 C_2 \beta_{12} 
\\
&\hphantom{+}+
C_1 C_3 \beta_{13} + C_1 C_2 \beta_{21} + C_2^2 \beta_{22} + 
C_2 C_3 \beta_{23} + C_1 C_3 \beta_{31} + C_2 C_3 \beta_{32} + 
C_3^2 \beta_{33} ]
\\
\mu_4 &= 
2 [ 3 C_2^2 + 3 A_0 (C_2 - C_3) + 12 C_2 C_3 + 3 C_3^2 ].
\end{align*}

%\subsection{$\xi_i$ and $\rho_i$}\label{xirho}
 
\section{Positivity conditions for quartics}
\label{appB}

For a quartic polynomial $ax^4 + bx^3+cx^2 + d x+e$, we define the discriminant $\Delta$, and the polynomials $P$ and $D$:
\begin{align}
\Delta &\equiv 256 a^3 e^3-192 a^2 b d e^2-128 a^2 c^2 e^2+144 a^2 c d^2 e-27 a^2 d^4 +144 a b^2 c e^2-6 a b^2 d^2 e-80 a b c^2 d e
\\
&\hphantom{=}\: \, +18 a b c d^3-4 a c^3 d^2+16 a c^4 e
-4 b^2 c^3 e+b^2 c d^2+18 b^3 c d e-4 b^3 d^3-27 b^4 e^2
\\
P&\equiv8 a c-3 b^2
\\
D&\equiv-16 a^2 b d-16 a^2 c^2+64 a^3 e+116 a b^2 c-3 b^4.
\end{align}
For the polynomial to always be positive, it must have $a>0$ and also have no real roots. For the latter to occur, it turns out that we must have $\Delta>0$, as well as \emph{either} $P>0$ or $D>0$.

\bibliographystyle{unsrtnat}
\bibliography{refs}

\end{document}